\documentclass[floatfix,aps,prd,reprint,nofootinbib,superscriptaddress,amsmath,amssymb]{revtex4-2}

\usepackage[italicdiff]{physics}
\usepackage{blindtext}
\usepackage{graphicx}
\usepackage{dcolumn}
\usepackage{hyperref}
\usepackage{bm}
\usepackage{color}

\def\eq{Eq.}
\def\eqs{Eqs.}

\begin{document}

\title{Bona-Mass\'o slices of Reissner-Nordstr\"om spacetimes}
\author{Sean E. Li} 
\email{sli@bowdoin.edu}
\affiliation{Department of Physics and Astronomy, Bowdoin College, Brunswick, Maine 04011, USA}
\author{Thomas W. Baumgarte}
\email{tbaumgar@bowdoin.edu}
\affiliation{Department of Physics and Astronomy, Bowdoin College, Brunswick, Maine 04011, USA}
\author{Kenneth A. Dennison}
\email{kdenniso@bowdoin.edu}
\affiliation{Department of Physics and Astronomy, Bowdoin College, Brunswick, Maine 04011, USA}
\author{H. P. de Oliveira}
\email{henrique.oliveira@uerj.br}
\affiliation{Department of Physics and Astronomy, Bowdoin College, Brunswick, Maine 04011, USA}
\affiliation{Departamento de F\'isica Te\'orica, Instituto de F\'isica A. D. Tavares, Universidade do Estado do Rio de Janeiro, R. S\~ao Francisco Xavier, 524, 20550-013 Rio de Janeiro, Brazil}

\begin{abstract}
Motivated by recent numerical relativity simulations of charged black holes and their interactions, we explore the properties of common slicing conditions in Reissner-Nordstr\"om spacetimes.  Specifically, we consider different choices for the so-called Bona-Mass\'o function and construct static and spherically symmetric slices of the Reissner-Nordstr\"om spacetime satisfying the corresponding slicing conditions.  For some of these functions the construction is entirely analytical, while for others we use numerical root-finding to solve quartic equations.  Our solutions are parameterized by the charge-to-mass ratio $\lambda = Q/M$ and approach a unique slice, independent of the Bona-Mass\'o functions considered here, in the extremal limit $\lambda \to 1$.
\end{abstract}
\maketitle

\section{\label{sec:intro}Introduction}

In many numerical relativity simulations, the time coordinate is specified by imposing a slicing condition for the lapse function $\alpha$.  A very common condition is the Bona-Mass\'o slicing condition \cite{BonaMasso1994}
\begin{eqnarray}
    \qty(\partial_t - \beta^i \partial_i)\alpha = -\alpha^2 f(\alpha) K, \label{eq:bona-masso-condition}
\end{eqnarray}
where $\beta^i$ is the shift vector, $f(\alpha)$ a yet-to-be-specified function of the lapse, and $K \equiv K^i_{~i}$ the mean curvature, i.e.~the trace of the extrinsic curvature. Choosing a Bona-Mass\'{o} function $f(\alpha)$ identifies a specific slicing of the spacetime; for $f = 1$, for example, (\ref{eq:bona-masso-condition}) reduces to harmonic slicing.  A very common choice is $f = 2/\alpha$, which results in so-called ``1+log" slicing.  Combined with a ``Gamma-driver" condition for the shift \cite{AlcB01,AlcBDKPST03,vanMBKC06}, 1+log slicing forms the so-called moving-puncture coordinates that have been used, for example, in numerous simulations of black-hole binaries (see, e.g., \cite{CamLMZ06,BakCCKM06}).

Significant insight into the properties of 1+log slicing, and hence our understanding of the above simulations, resulted from analytical studies of 1+log slices of the Schwarzschild spacetime (e.g., \cite{HanHPBM07,HanHBGSO07,BauN07,HanHOBO08,Brugmann2009}). In particular, these studies revealed the trumpet geometry of the resulting slices, which helped to explain their remarkable numerical properties.  

In recent years, several authors have also considered black holes with charge, and have simulated their interaction in the framework of Einstein-Maxwell theory \cite{Alcubierre2009,Zilhao2012,Zilhao2013,Zilhao2014, Zilhao2015,Lehner2017,BozzolaPaschalidis2021,Bozzola2021,Bozzola2022,Mukherjee2022,LunaBozzola2022}. In part, these simulations are motivated by astrophysical considerations---for example, to explore whether current observations of gravitational-wave signals can be used to place bounds on the black-hole charge---and in part by the recognition that Einstein-Maxwell theory is a well-posed example of a tensor-vector theory, and may therefore serve as a stand-in for more exotic extensions of general relativity. Many of the above simulations also adopt 1+log slicing, raising the question of whether its desirable properties for uncharged black holes also exist for charged black holes.

Motivated by these considerations we generalize in this paper previous work on Bona-Mass\'o slices of Schwarzschild spacetimes to their charged counterparts, namely Reissner-Nordstr\"om (RN) spacetimes. Specifically, we follow \cite{BaumgarteOliveira2022} and consider a number of different families of Bona-Mass\'o functions $f(\alpha)$, but apply these to charged, rather than uncharged, static black holes. We outline our mathematical approach in Section \ref{sec:methods}, consider extremal black holes in Section \ref{sec:extremal}, discuss results for specific choices of the Bona-Mass\'o function in Section \ref{sec:slices}, and briefly summarize in Section \ref{sec:summary}.

Throughout this paper we use geometrized units with $G = c = 1$ and adopt the convention that indices $a, b, c, \ldots$ represent spacetime indices while $i, j, k, \ldots$ denote spatial indices.

\section{\label{sec:methods} Basic equations}

Most of this section is a direct extension of previous work on 1+log slices of Schwarzschild spacetimes, e.g., \cite{HanHPBM07,HanHOBO08,Brugmann2009}.  We generalize those previous treatments by considering different families of Bona-Mass\'o functions (see also \cite{BaumgarteOliveira2022}) and by applying these to Reissner-Nordstr\"om spacetimes.

\subsection{\label{subsec:integral}Transformation to Bona-Mass\'o slices}

The line element for a non-rotating, charged black hole can be written as 
\begin{eqnarray}
    \dd{s^2} &= -f_0 \dd{t^2} + f_0^{-1} \dd{R^2} + R^2 \dd{\Omega^2}, \label{eq:reissner-nordstrom-metric}
\end{eqnarray}
where we have defined
\begin{equation}
    f_0 \equiv 1 - \dfrac{2M}{R} + \dfrac{Q^2}{R^2}  \label{eq:f-naught}
\end{equation}
(not to be confused with the Bona-Mass\'o function $f(\alpha)$ defined in \eq~(\ref{eq:bona-masso-condition})).  In the above equations, $R$ is the areal radius, $M$ the black-hole mass, and $Q$ the black-hole charge. We also note that the two horizons of an RN spacetime are located at the roots of the function $f_0 = f_0(R)$, i.e.~at
\begin{equation} \label{eq:horizons}
    R_\pm = M \pm \sqrt{M^2 - Q^2}.
\end{equation}

We now transform to new spatial slices using a height-function approach (see, e.g., Sec.\ IV.2 in \cite{BaumgarteShapiro2010} for a textbook treatment), i.e.~we write a new time coordinate $\bar t$ as
\begin{eqnarray}
    \bar{t} &= t + h(R).  \label{eq:t-bar}
\end{eqnarray}
By allowing the height function $h = h(R)$ to depend on radius only, we restrict our focus to time-independent and spherically symmetric slices.  Inserting (\ref{eq:t-bar}) into the line element (\ref{eq:reissner-nordstrom-metric}) then yields\footnote{See also \cite{Reimann2004A, Reimann2004B}, who adopted the height-function approach to construct maximal slices in non-extremal RN spacetimes, and \cite{PanossoJaramillo2018}, who constructed hyperboloidal slices of RN spacetimes using this approach.} 
\begin{eqnarray}
    \dd{s^2} &=& -f_0\dd{\bar{t}}^2 + 2 f_0 h' \dd{\bar{t}}\dd{R} \nonumber \\
    & & + \qty(f_0^{-1} - f_0 {h'}^2)\dd{R}^2 + R^2\dd{\Omega}^2 \label{eq:new-line-element}
\end{eqnarray}
where the prime denotes differentiation with respect to $R$, $h' \equiv \dv*{h}{R}$. We compare (\ref{eq:new-line-element}) with the general 3+1 form of the spacetime metric, 
\begin{eqnarray}
    \dd{s}^2 &= -\alpha^2 \dd{\bar{t}}^2 + \gamma_{ij} \qty(\dd{x}^i + \beta^i \dd{\bar{t}})\qty(\dd{x}^j + \beta^j \dd{\bar{t}}),
\end{eqnarray}
to identify the $RR$-component of the spatial metric
\begin{eqnarray}
    \gamma_{RR} &= f_0^{-1} - f_0 {h'}^2, \label{eq:spatial-metric-RR}
\end{eqnarray}
the $R$-component of the shift vector
\begin{eqnarray}
    \beta^R = \dfrac{f_0 h'}{\gamma_{RR}} = \dfrac{f_0^2 h'}{1 - f_0^2 {h'}^2}, \label{eq:shift-R}
\end{eqnarray}
and the square of the lapse 
\begin{eqnarray}
    \alpha^2 = f_0 + \gamma_{RR}\qty(\beta^R)^2 = \dfrac{f_0}{1 - f_0^2 {h'}^2}. \label{eq:lapse-h}
\end{eqnarray}
We note that $\alpha$ does not necessarily vanish at a root of $f_0$, i.e.~on the black-hole horizons, since $h'$ may diverge there.  Using (\ref{eq:lapse-h}) we may also rewrite the shift (\ref{eq:shift-R}) as 
\begin{eqnarray}
    \beta^R = \alpha\sqrt{\alpha^2 - f_0} = \alpha^2 | f_0 {h'} | \label{eq:shift-R-no-h}
\end{eqnarray}
where we have taken a positive root.  We compute the mean curvature from
\begin{eqnarray}
    K = -\nabla_a n^a = -\dfrac{1}{\sqrt{\lvert g \rvert}}\partial_a\qty(\sqrt{\lvert g \rvert} \, n^a), \label{eq:mean-curvature-identity}
\end{eqnarray}
where $g = - R^4 \sin^2 \theta$ is the determinant of the metric, and $n^a$ the future-oriented normal of the hypersurface
\begin{eqnarray}
    n^a = \alpha^{-1} \qty(1, -\beta^i). \label{eq:normal-vector}
\end{eqnarray}
For static and spherically symmetric slices, (\ref{eq:mean-curvature-identity}) becomes
\begin{eqnarray}
    K &=& \dfrac{1}{R^2}\frac{d}{dR} \qty(R^2 \dfrac{\beta^R}{\alpha}) 
    = \dfrac{2}{R}\dfrac{\beta^R}{\alpha} + \dfrac{\qty(\beta^R)'}{\alpha} - \dfrac{\beta^R}{\alpha^2}\alpha' \label{eq:mean-curvature}
\end{eqnarray}
and the Bona-Mass\'o condition (\ref{eq:bona-masso-condition}) reduces to
\begin{eqnarray}
    \beta^R \alpha' = \alpha^2 f(\alpha) K. \label{eq:bona-masso-reduced}
\end{eqnarray}
Substituting (\ref{eq:mean-curvature}) into (\ref{eq:bona-masso-reduced}) then yields
\begin{eqnarray}
    \dfrac{\dd{\alpha}}{\alpha f(\alpha)} + \dfrac{\dd{\alpha}}{\alpha} = \dfrac{2\dd{R}}{R} + \dfrac{\dd{\beta^R}}{\beta^R}, \label{eq:bona-masso-integrand}
\end{eqnarray}
which, using (\ref{eq:shift-R-no-h}), we may integrate to obtain
\begin{eqnarray}
    \alpha^2 &=& 1 - \dfrac{2M}{R} +\dfrac{Q^2}{R^2} + \dfrac{C e^{2I(\alpha)}}{R^4} \nonumber \\
    &=& f_0(R) + \dfrac{C e^{2I(\alpha)}}{R^4}. \label{eq:first-integral}
\end{eqnarray}
In (\ref{eq:first-integral}) we defined the integral
\begin{eqnarray}
    I(\alpha) \equiv \int_0^\alpha \dfrac{\dd{\Tilde{\alpha}}}{\Tilde{\alpha} f(\Tilde{\alpha})}, \label{eq:integral-of-alpha}
\end{eqnarray}
and $C$ is an undetermined constant of integration with units of $M^4$.  We note that the above expressions differ from their counterparts for uncharged black holes only by the appearance of the term $Q^2/R^2$ in \eq~(\ref{eq:first-integral}).


\subsection{\label{subsec:regularity}Regularity condition}

We will be interested in regular slices that penetrate the outer horizon $R_+$ (see (\ref{eq:horizons})), meaning that the lapse $\alpha$ should connect $\alpha = 1$ in the asymptotic region $R \rightarrow \infty$ with a root $\alpha = 0$ at a radius $R_0 \leq R_+$.  If such a slice penetrated the inner horizon also we would have $f_0(R_0) > 0$, in which case we could evaluate (\ref{eq:first-integral}) at $R_0$ to find $C < 0$.  At either one of the horizons, however, (\ref{eq:first-integral}) would then yield $\alpha^2 < 0$, which does not have a real solution. We therefore conclude that regular slices can penetrate the outer horizon only; those that do penetrate the outer horizon then have $R_- \leq R_0 \leq R_+$ with $f_0(R_0) \leq 0$ and hence $C \geq 0$. As we observed below \eq~(\ref{eq:lapse-h}), the derivative of the height function $h'$ will necessarily diverge at $R_+$ for such a slice.

For general values of the constant $C$ in \eq~(\ref{eq:first-integral}) the resulting lapse $\alpha$ will {\em not} connect a root at $R_0$ with the asymptotic region; instead, there may be regions at radii $R > R_0$ for which (\ref{eq:first-integral}) does not yield real values of $\alpha$ at all.  The regular slices that we are interested in therefore exist for special values of $C$ only.  In order to identify these values of $C$ we follow \cite{HanHBGSO07,HanHOBO08} and consider an equation for the derivative of the lapse. Inserting (\ref{eq:shift-R-no-h}) into both (\ref{eq:mean-curvature}) and (\ref{eq:bona-masso-reduced}) we obtain
\begin{eqnarray}
    \alpha' & = & \dfrac{\alpha f(\alpha)}{M\hat{R}}\dfrac{2 - 3/\hat{R} + \lambda^2/\hat{R}^2 - 2\alpha^2}{1 - 2/\hat{R} + \lambda^2/\hat{R}^2 + \alpha^2 f(\alpha) - \alpha^2} \nonumber \\
    & = & \dfrac{\alpha f(\alpha)}{M\hat{R}}\dfrac{2 - 3/\hat{R} + \lambda^2/\hat{R}^2 - 2\alpha^2}{f_0(R) + \alpha^2 (f(\alpha) - 1)}
    \label{eq:lapse-derivative}
\end{eqnarray}
where we have introduced a dimensionless areal radius $\hat{R} \equiv R/M$ and the dimensionless charge-to-mass ratio $\lambda \equiv Q/M$. We now observe that the denominator on the right-hand side may have a root for $R > R_0$; if so, $\alpha'$ can remain regular at that root of the denominator only if the numerator has a simultaneous root. The radius and lapse at such a critical point (denoted by $\hat R_c$ and $\alpha_c$) must therefore satisfy the two equations\footnote{Assuming that $\alpha^2(f(\alpha) - 1)$ vanishes for $\alpha = 0$, and that $f(\alpha) > 1$ for all $\alpha$, horizon-penetrating slices necessarily go through a critical point at a point $R_c$ such that $R_0 \leq R_c \leq R_+$. This is because $f_0(R)$ has a root at $R_+$, while $-\alpha^2(f(\alpha) - 1)$ has a root at $R_0$; both are non-positive between these two points, and intersect so that the denominator of the second term of (\ref{eq:lapse-derivative}) vanishes. The assumption $f(\alpha) > 1$ for all $\alpha$ holds for most Bona-Mass\'o functions considered in this paper, but not for the analytical trumpet slices of Section \ref{subsec:trumpet}.   For the latter it is possible to construct slices that avoid a critical point altogether, but we will instead focus on slices that pass through a critical point in this paper.}
\begin{subequations} \label{eq:zero}
\begin{align}
    2 - \dfrac{3}{\hat{R}_c} + \dfrac{\lambda^2}{\hat{R}_c^2} - 2\alpha_c^2 &= 0 \quad\text{and} \label{eq:numerator-zero} \\
    1 - \dfrac{2}{\hat{R}_c} + \dfrac{\lambda^2}{\hat{R}_c^2} + \alpha_c^2 f(\alpha_c) - \alpha_c^2 &= 0, \label{eq:denominator-zero}
\end{align}
\end{subequations}
where we have assumed that the root of the denominator of (\ref{eq:lapse-derivative}) results from a vanishing of the denominator of the second fraction in (\ref{eq:lapse-derivative}), rather than the first.

We can eliminate $\lambda$ from \eqs~(\ref{eq:zero}) to obtain one equation for $\hat R_c$ and $\alpha_c$ alone,
\begin{eqnarray}
    \hat{R}_c = \dfrac{1}{1 - \alpha_c^2 f(\alpha_c) - \alpha_c^2}. \label{eq:critical-point-relation}
\end{eqnarray}
We then re-insert (\ref{eq:critical-point-relation}) into (\ref{eq:numerator-zero}) and, depending on the specific choice of $f(\alpha)$, find $\alpha_c$ either by numerical root-finding for a given $\lambda$ or by solving for $\alpha_c$ analytically. Given $\alpha_c$ we then find $\hat R_c$ from (\ref{eq:critical-point-relation}), and finally insert both into (\ref{eq:first-integral}) to obtain the constant $C$.

The above procedure works as long as $\alpha f(\alpha)$ in the first term on the right-hand side of (\ref{eq:lapse-derivative}) remains finite as $\alpha \to 0$.  This is the case for most Bona-Mass\'o functions considered in this paper, but not for the shock-avoiding slices with $f(\alpha) = 1 + \kappa / \alpha^2$ (see \cite{Alcubierre1997}).  For the latter, the (outer-most) root of the denominator of \eq~(\ref{eq:lapse-derivative}) occurs for $\alpha = 0$ rather than a root of (\ref{eq:denominator-zero}), provided $\kappa$ satisfies condition (\ref{eq:second-kappa-condition}).  Inserting $\alpha_c = 0$ into (\ref{eq:numerator-zero}) then yields the critical radius $\hat{R}_c$ (see also Section \ref{subsubsec:full} below).

\subsection{\label{subsec:root}The root of the lapse}

For a given $f(\alpha)$ whose integral $I(\alpha)$ is known, evaluating (\ref{eq:first-integral}) at the critical point, for known values of $\hat R_c$ and $\alpha_c$, allows computing the constant of integration $C$. From there, we compute the root $\hat{R}_0$ of the lapse by setting $\alpha = 0$ in (\ref{eq:first-integral}).  For some choices of the Bona-Mass\'o function $f(\alpha)$ we can solve the resulting quartic equation for $\hat{R}_0$ analytically, while for others we use numerical root-finding.  

As the next step we compute the dimensionless derivative of the lapse evaluated at its root,
\begin{equation}
    \hat{a}_1 \equiv M \alpha'(\hat{R}_0).
\end{equation}
In most cases this is done via implicit differentiation of (\ref{eq:first-integral}), except for the ``fully shock-avoiding" slicing condition of Section \ref{subsec:noshock}, for which it is more convenient to take the limit $\hat{R} \to \hat{R}_0$ of \eq~(\ref{eq:lapse-derivative}). In order to choose valid roots $\hat{R}_0$ among the real solutions to the above quartic equation we check that $\hat{a}_1 \geq 0$ so that the lapse stays non-negative near the root.

\subsection{\label{subsec:isotropic}Transformation to isotropic coordinates}

Finally, we transform to isotropic coordinates with radial coordinate $r$.  To do so, we compare the spatial part of the line element (\ref{eq:new-line-element}),
\begin{eqnarray}
    \dd{l}^2 = \alpha^{-2} \dd{R}^2 + R^2\dd{\Omega}^2, \label{eq:spatial-line-element}
\end{eqnarray}
with the spatial line element in isotropic coordinates,
\begin{eqnarray}
    \dd{l}^2 = \psi^4 \qty(\dd{r}^2 + r^2\dd{\Omega}^2), \label{eq:spatial-line-element-isotropic}
\end{eqnarray}
where $\psi$ is a conformal factor, which yields the system
\begin{subequations} \label{eq:conformal-condition}
\begin{align}
    \alpha^{-1}\dd{R} &= \psi^2 \dd{r} \qq{and} \label{eq:conformal-condition-one}\\
    R &= \psi^2 r. \label{eq:conformal-condition-two}
\end{align}
\end{subequations}
Together, \eqs~(\ref{eq:conformal-condition}) yield
\begin{eqnarray}
    \dfrac{\dd{r}}{r} = \dfrac{\dd{R}}{R\alpha} = \dfrac{\dv*{R}{\alpha}}{R}\dfrac{\dd{\alpha}}{\alpha}, \label{eq:isotropic-first-integral}
\end{eqnarray}
which we may integrate to obtain
\begin{eqnarray}
    r = \exp \int \dfrac{\dv*{R}{\alpha}}{R}\dfrac{\dd{\alpha}}{\alpha}. \label{eq:areal-isotropic-relation}
\end{eqnarray}
Making the leading-order approximation $\dv*{R}{\alpha} \simeq 1/a_1$ and $R \simeq R_0$, we further integrate (\ref{eq:areal-isotropic-relation}) to find
\begin{eqnarray} \label{eq:power_law}
    \alpha \propto r^{1/\gamma} \quad (r \to 0),
\end{eqnarray}
where we have adopted the notation of \cite{Brugmann2009} in defining 
\begin{eqnarray}
    \gamma \equiv \frac{1}{\hat{a}_1 \hat{R}_0}.
\end{eqnarray}

For general values of $R > R_0$ we integrate (\ref{eq:areal-isotropic-relation}) following the prescription laid out in \eqs~(46), (47) and (67) of \cite{Brugmann2009} in order to obtain the isotropic radius $r$ as a function of the areal radius $R$.  As a consistency check we verify that, in the vicinity of the root, the lapse behaves according to the power law (\ref{eq:power_law}) (see also \eq~(56) in \cite{Brugmann2009}). 
Having obtained $r$, we can compute the conformal factor $\psi$ from (\ref{eq:conformal-condition-two}) as
\begin{eqnarray}
    \psi = \sqrt{\dfrac{R}{r}} \label{eq:conformal-factor},
\end{eqnarray}
where we see that near the root of the lapse, where $R$ approaches $R_0$, we have 
\begin{eqnarray}
    \psi \propto r^{-1/2}  \quad (r \to 0) \label{eq:conformal-factor-near-root}
\end{eqnarray}
as is characteristic for a trumpet geometry.

\section{\label{sec:extremal}The extremal limit}

Before discussing specific choices for the Bona-Mass\'o function $f(\alpha)$ in Section \ref{sec:slices} we first consider extremal Reissner-Nordstr\"om black holes with $Q = M$, i.e.~$\lambda = 1$.

For $\lambda = 1$, \eq~(\ref{eq:first-integral}) becomes
\begin{eqnarray}
    \alpha^2 = \qty(\dfrac{\hat{R} - 1}{\hat{R}})^2 + \dfrac{\hat{C} e^{2I(\alpha)}}{\hat{R}^4},
\end{eqnarray}
where we have defined the dimensionless constant of integration $\hat{C} = M^{-4} C$. Since the exponential term is always positive, solutions for the lapse must have $C \leq 0$ in order to have a root in this case, independently of the choice of $f(\alpha)$.

As we discussed in Section \ref{subsec:regularity}, the procedure for finding the critical point depends on the behavior of $\alpha f(\alpha)$ as $\alpha \to 0$.  If $\alpha f(\alpha)$ remains finite in this limit, we identify the critical point by finding simultaneous roots of \eqs~(\ref{eq:zero}).  In the extremal limit, we may then rewrite \eq~(\ref{eq:denominator-zero}) as
\begin{eqnarray}
    \qty(1 - \dfrac{1}{\hat{R}_c})^2 + \alpha_c^2 \left( f(\alpha_c) - 1 \right) = 0. \label{eq:extremal-denominator-zero}
\end{eqnarray}
Assuming $f(\alpha_c) > 1$, (\ref{eq:extremal-denominator-zero}) implies that the only critical point for non-negative $\alpha_c$ occurs at $\hat{R}_c = 1$ with $\alpha_c = 0$ (for which (\ref{eq:numerator-zero}) features a root also). Inserting these values into (\ref{eq:first-integral}) then yields $C = 0$ in the extremal limit.

For shock-avoiding slices with $f(\alpha) = 1 + \kappa / \alpha^2$, on the other hand, the critical point is given by $\alpha_c = 0$ and $\hat{R}_c$ by a root of (\ref{eq:numerator-zero}).  In the extremal limit, these two roots are $\hat R_0 = 1$ and $\hat R_0 = 1/2$.  Only for the former, however, does $\hat {a}_1$ take a non-imaginary value, so that we obtain the exact same critical values as in the case above.

For all Bona-Mass\'o functions $f(\alpha)$ considered here we therefore have $\hat{C} = 0$ in the extremal limit, so that \eq~(\ref{eq:first-integral}) yields
\begin{equation} \label{eq:alpha_of_R_extremal}
    \alpha = \frac{\hat R - 1}{\hat R}
\end{equation}
independently of $f(\alpha)$. Finally, we observe that we have $\hat a_1 = 1$ and hence $1/\gamma = 1$ in the extremal limit.

\section{\label{sec:slices}Results for specific slices}
\begin{figure}[t]
    \centering
    \includegraphics[width=\linewidth]{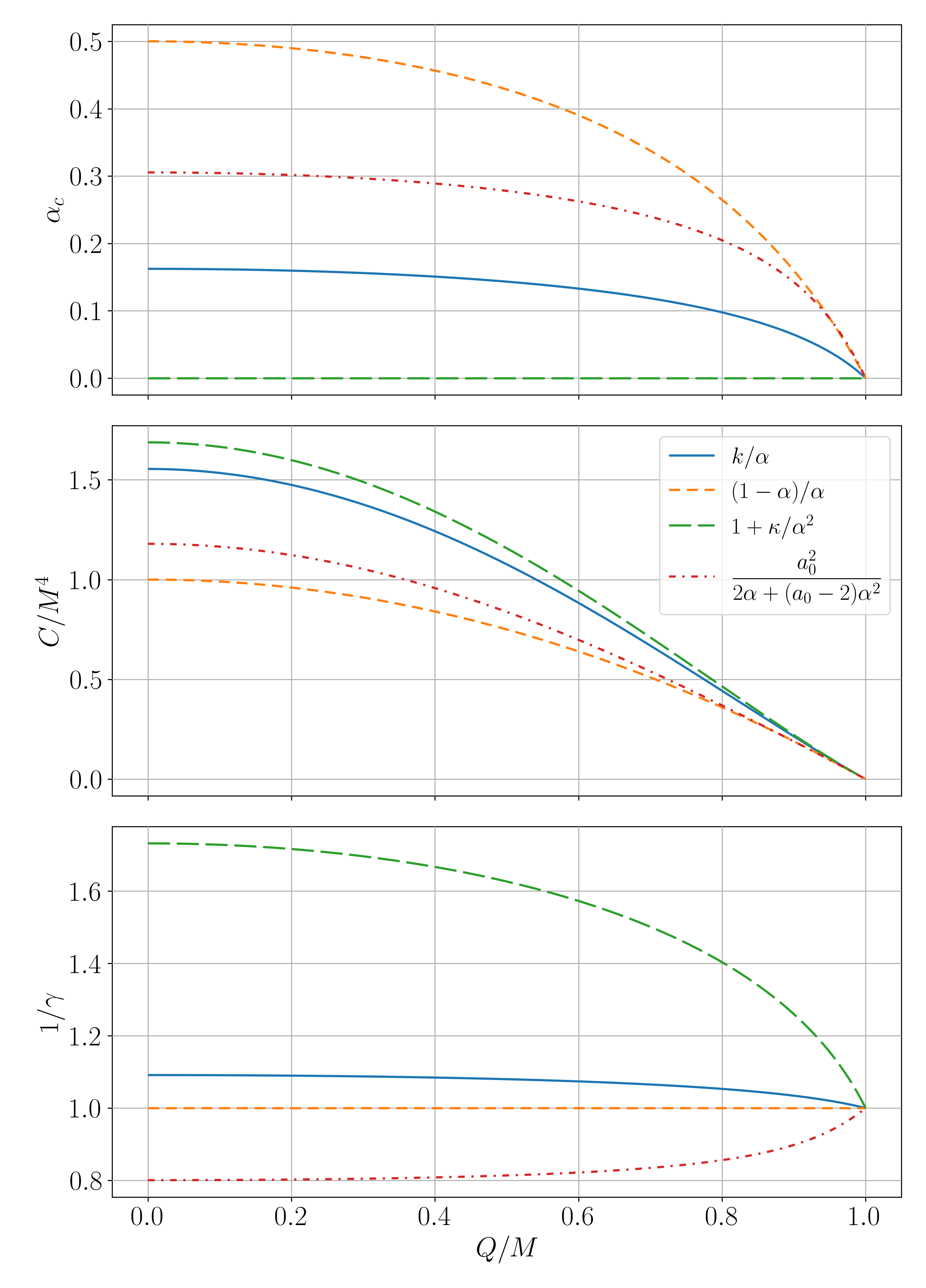}
    \caption{The critical lapse $\alpha_c$ (top), constant of integration $C$ (center), and exponent $1/\gamma$ (bottom) plotted against the charge-to-mass ratio $\lambda$ (up to 0.999) on a shared horizontal axis for each of the slices given by $f(\alpha)$ that we consider. The solid (blue) line corresponds to 1+log slices (with $k = 2$), the short-dashed (orange) line to analytical trumpet slices, the long-dashed (green) line to fully gauge-shock-avoiding slices (with $\kappa = 1$), and the dash-dotted (red) line to zeroth-order gauge-shock-avoiding slices (with $a_0 = 4/3$). All slicing conditions yield the same critical values in the extremal limit $Q \to M$.}
    \label{fig:all-alpha_c-C-gamma}
\end{figure}
\begin{figure}[t]
    \centering
    \includegraphics[width=\linewidth]{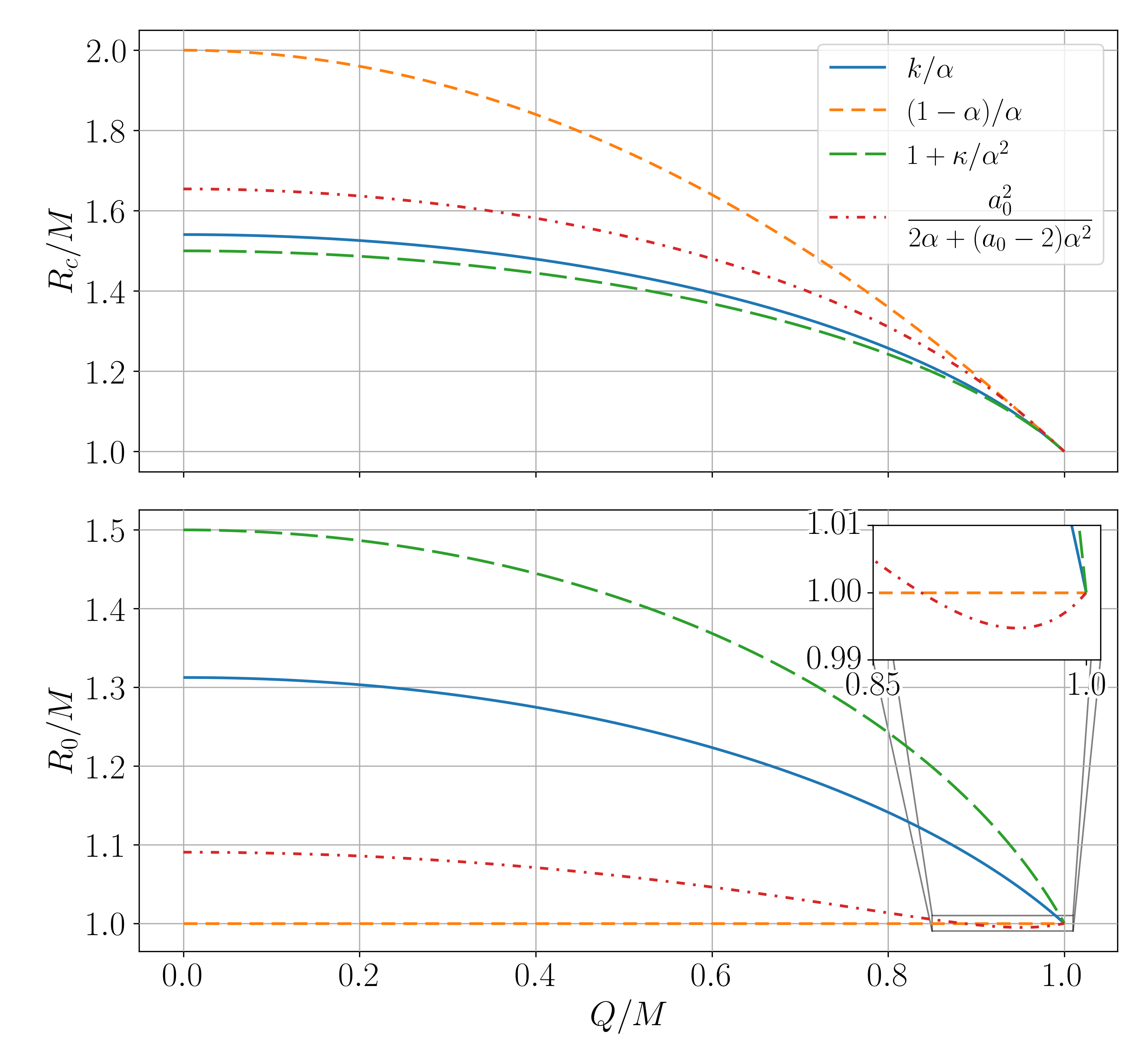}
    \caption{The critical areal radius $R_c$ (top) and the areal radius $R_0$ at which the lapse vanishes (bottom) versus $\lambda$ for each $f(\alpha)$. The inset in the top-right corner of the bottom panel shows an expanded view of the bottom-right region, where the root of the lapse for zeroth-order shock-avoiding slices falls slightly below $\hat{R}_0 = 1$ near the extremal limit.}
    \label{fig:all-R_c-R_0}
\end{figure}
\begin{figure}[t]
    \centering
    \includegraphics[width=\linewidth]{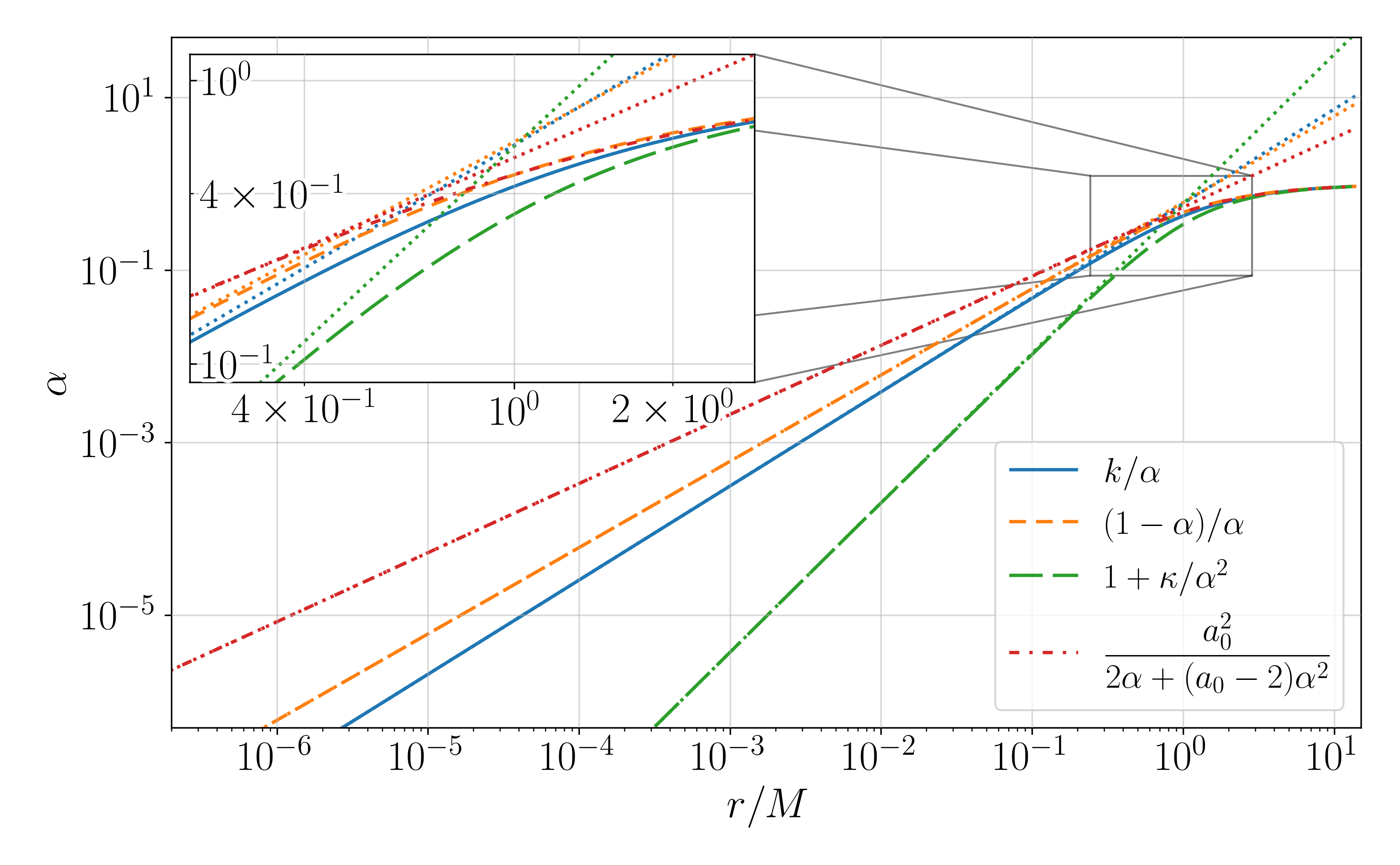}
    \caption{Profiles of the lapse $\alpha$ as a function of isotropic radius $r$ for each of the slices we consider (with $k = 2$, $\kappa = 1$, and $a_0 = 4/3$), with charge-to-mass ratio $\lambda = 0$ (the Schwarzschild spacetime). The dotted lines represent the expected power-law behavior $\alpha \propto r^{1/\gamma}$ in the limit $r \to 0$. The inset in the top-left corner expands a crowded region of the plot where the lapse profiles depart from their small-radius power-law behavior.}
    \label{fig:all-lapse-isotropic-zero-charge}
\end{figure}
\begin{figure}[t]
    \centering
    \includegraphics[width=\linewidth]{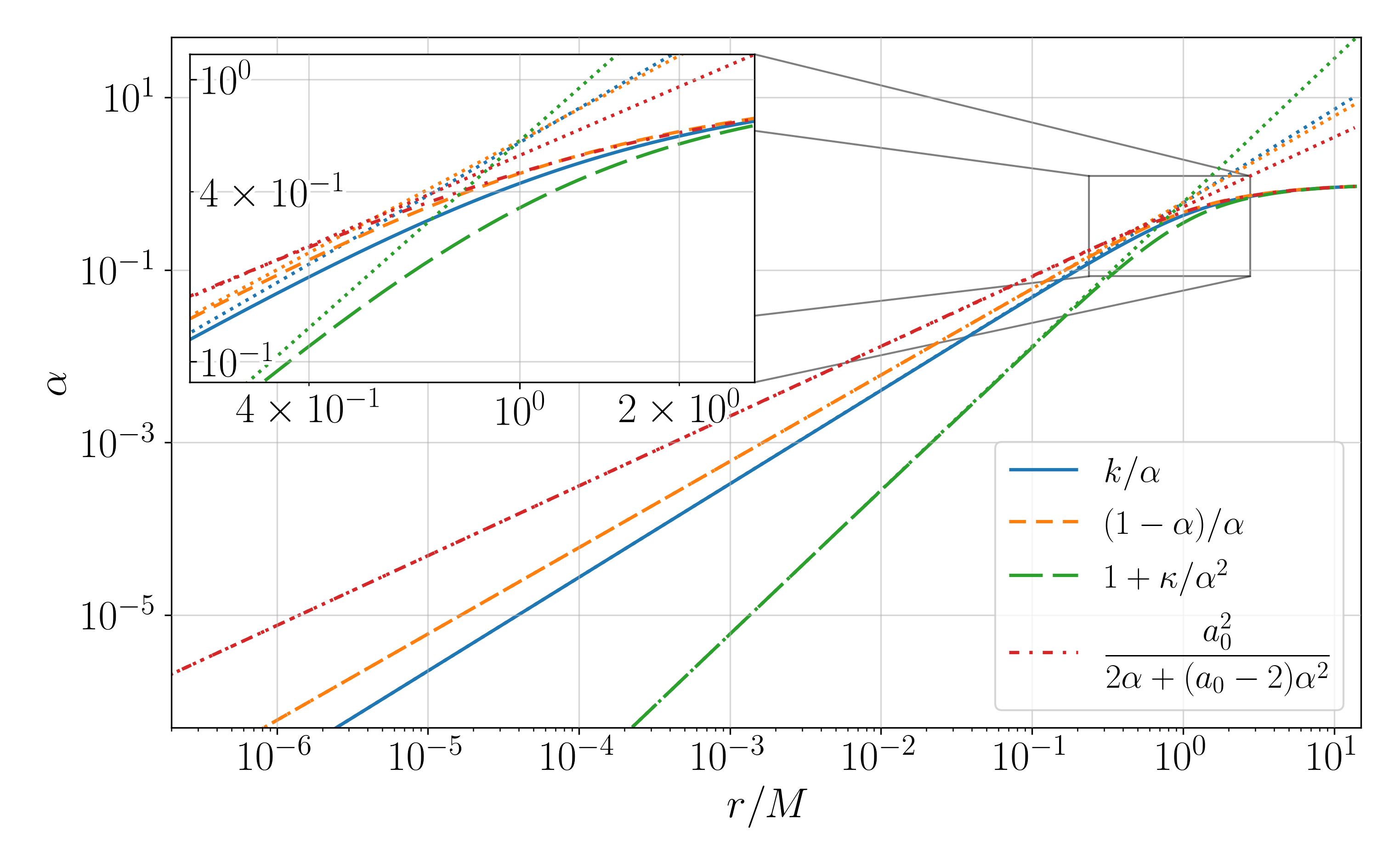}
    \caption{Same as Fig.~\ref{fig:all-lapse-isotropic-zero-charge}, but with $\lambda = 0.400$.}
    \label{fig:all-lapse-isotropic-q-equals-point-four}
\end{figure}
\begin{figure}[t]
    \centering
    \includegraphics[width=\linewidth]{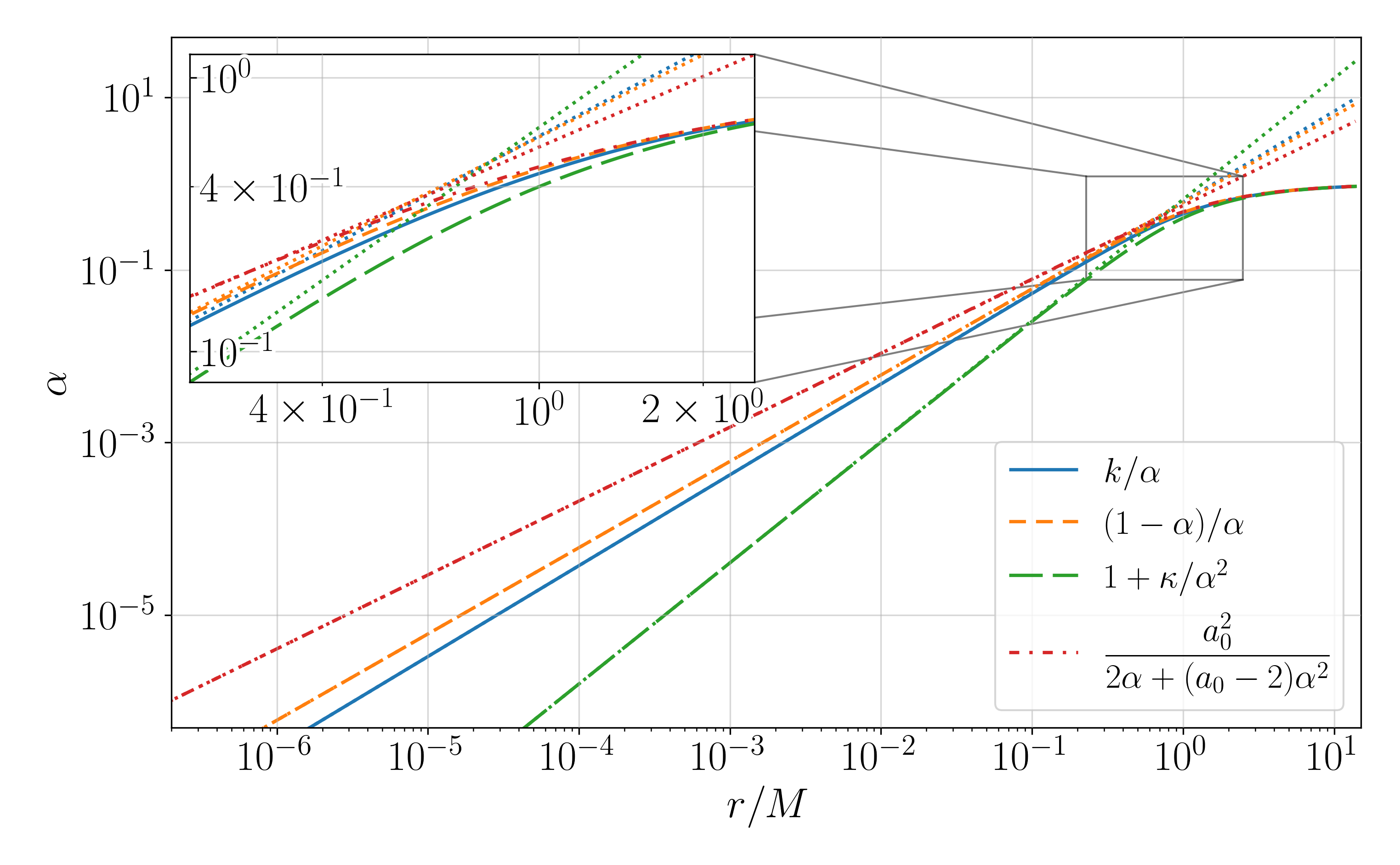}
    \caption{Same as Fig.~\ref{fig:all-lapse-isotropic-zero-charge}, but with $\lambda = 0.800$.}
    \label{fig:all-lapse-isotropic-q-equals-point-eight}
\end{figure}
\begin{figure}[t]
    \centering
    \includegraphics[width=\linewidth]{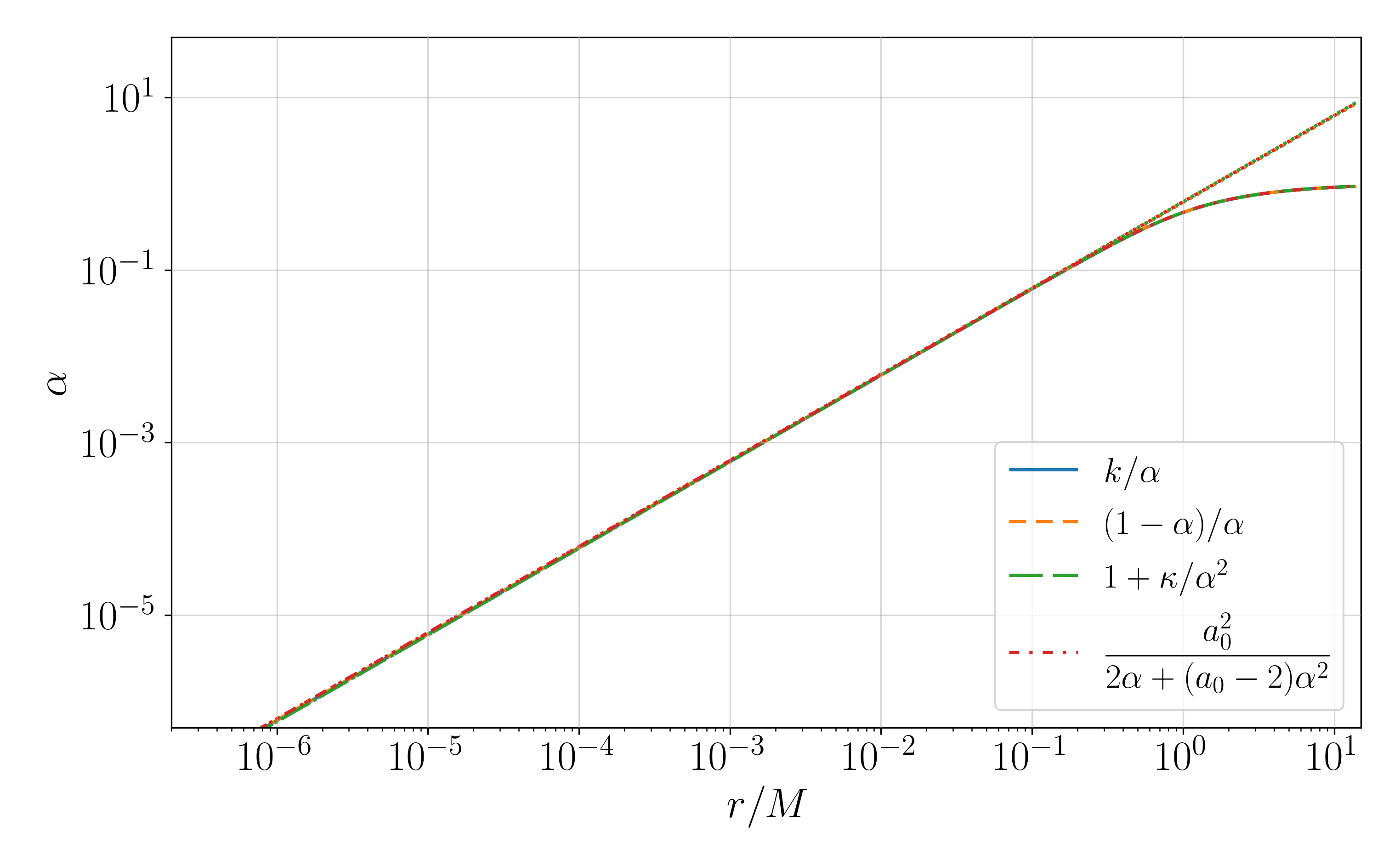}
    \caption{Same as Fig.~\ref{fig:all-lapse-isotropic-zero-charge}, but with $\lambda = 0.999$.}
    \label{fig:all-lapse-isotropic-extremal}
\end{figure}

In the following we consider four different families of Bona-Mass\'o functions and explore the associated slicing conditions. For each one we compute values of the parameters $\hat{R}_c$, $\alpha_c$, $\hat{C}$, $\hat{R}_0$, and $1/\gamma$ for different values of $\lambda = Q/M$ (see Figs.~\ref{fig:all-alpha_c-C-gamma} and \ref{fig:all-R_c-R_0}). For each family we also compute profiles of the lapse\footnote{\label{footnote:root-finding}
The lapse $\alpha$ as a function of $R$ can be found from \eq~(\ref{eq:first-integral}) using root-finding; we found it helpful to adopt $\alpha_c$ as an initial guess.}
as a function of the isotropic radius and show results for selected values of $\lambda$ in Figs.~\ref{fig:all-lapse-isotropic-zero-charge} through \ref{fig:all-lapse-isotropic-extremal}.

\subsection{\label{subsec:onepluslog}1+log slicing}
\begin{figure}[t]
    \centering
    \includegraphics[width=\linewidth]{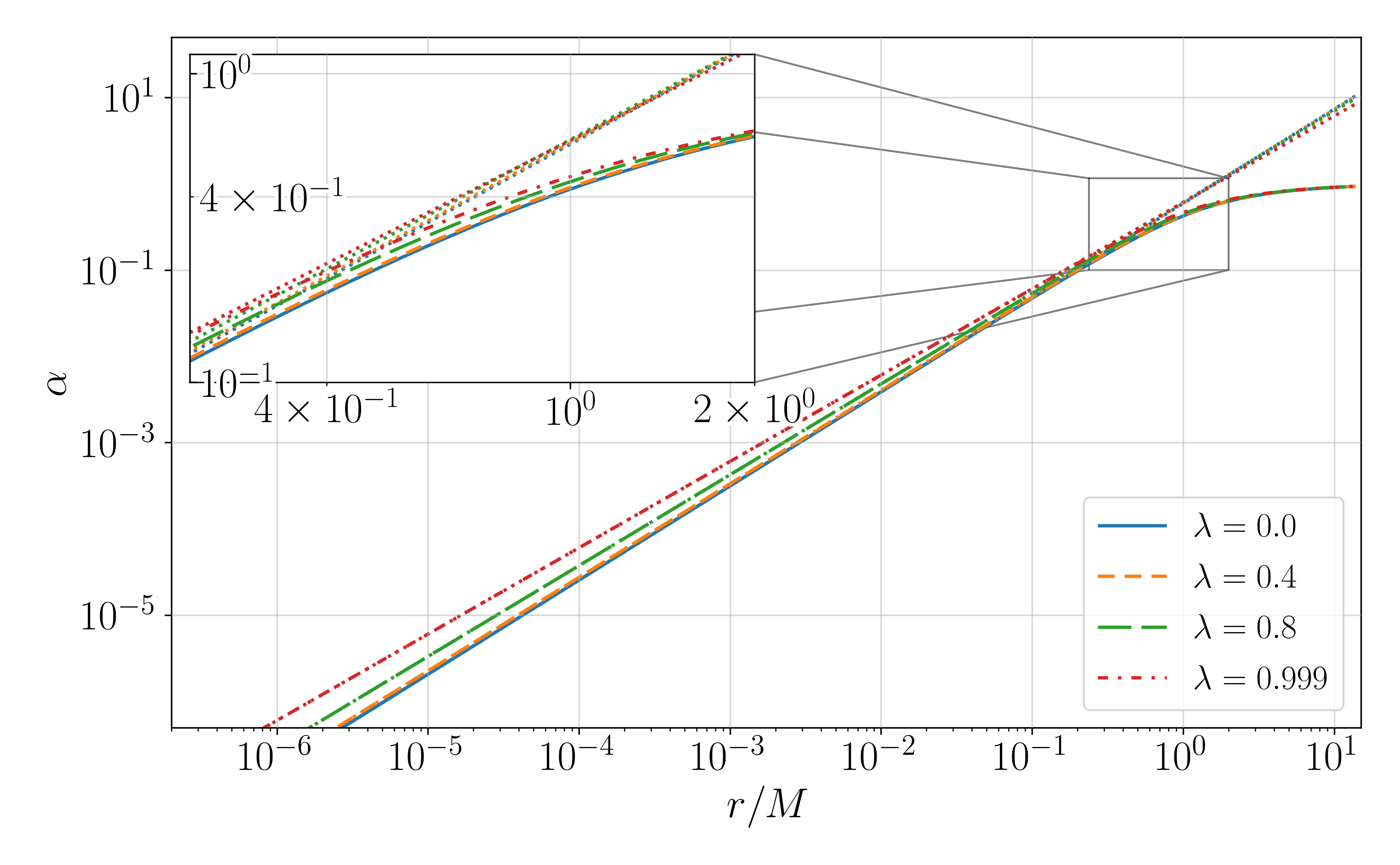}
    \caption{The numerical profile of the lapse $\alpha(r)$ for $\lambda \in \qty{0, 0.4, 0.8, 0.999}$ for 1+log slices with $k = 2$. See Fig.~\ref{fig:all-lapse-isotropic-zero-charge} for an explanation of the inset and dotted lines.}
    \label{fig:onepluslog-lapse}
\end{figure}

We first consider Bona-Mass\'o functions of the form
\begin{eqnarray}
    f(\alpha) = \dfrac{k}{\alpha}. \label{eq:onepluslog}
\end{eqnarray}
Even though, strictly speaking, 1+log slicing corresponds to the case $k = 2$ only (see \cite{BonaMasso1994}), we refer to the entire family as ``1+log" slicing.  For (\ref{eq:onepluslog}), the integral (\ref{eq:integral-of-alpha}) can be evaluated to yield $I(\alpha) = \alpha / k$, so that (\ref{eq:first-integral}) becomes 
\begin{eqnarray}
    \alpha^2 = 1 - \dfrac{2}{\hat{R}} + \dfrac{\lambda^2}{\hat{R}^2} + \dfrac{\hat{C}e^{2\alpha / k}}{\hat{R}^4}. \label{eq:onepluslog-first-integral}
\end{eqnarray}
Evaluating this either at the critical point (for $\alpha_c$ and $\hat R_c$) or at the root of the lapse (for $\alpha = 0$ and $\hat R_0$) yields two different expressions for the constant of integration, namely
\begin{eqnarray}
    \hat{C} &=& \hat{R}_c^4 e^{- 2\alpha_c / k} \left(\alpha_c^2 - 1 + \dfrac{2}{\hat{R}_c} - \frac{\lambda^2}{\hat{R}_c^2}\right) \label{eq:onepluslog-C-critical} \\ 
    &=& -\hat{R}_0^4 + 2\hat{R}_0^3 - \lambda^2 \hat{R}_0^2. \label{eq:onepluslog-C-root}
\end{eqnarray}
Inserting (\ref{eq:onepluslog}) into (\ref{eq:critical-point-relation}) and then substituting (\ref{eq:critical-point-relation}) for $\hat{R}_c$ in (\ref{eq:numerator-zero}), we obtain a quartic equation for $\alpha_c$ whose analytical solution is unwieldy. We thus use numerical root-finding to determine the critical point for these slices. The solution to (\ref{eq:onepluslog-C-root}) for $\hat{R}_0$ is similarly unwieldy, so we again use numerical root-finding to locate the root of the lapse.  

Implicit differentiation of (\ref{eq:onepluslog-first-integral}) yields
\begin{eqnarray}
    \hat{a}_1 = \dfrac{-2\hat{R}_0^3 + 2\lambda^2\hat{R}_0^2 + 4\hat{C} }{(2\hat{C}/k)\hat{R}_0},
\end{eqnarray}
from which we evaluate the exponent $1/\gamma = \hat{a}_1 \hat{R}_0$.  We show graphs of all the above parameters, as a function of $\lambda$, in Figs.~\ref{fig:all-alpha_c-C-gamma} and \ref{fig:all-R_c-R_0}, together with the corresponding results for the other slicing conditions discussed in the following subsections.

Finally, we carry out the transformation from the areal radius $R$ to isotropic radius $r$ as discussed in Section \ref{subsec:isotropic}, and show profiles of the lapse for 1+log slices for a few selected values of $\lambda$ in Fig.~\ref{fig:onepluslog-lapse}.

\subsection{\label{subsec:trumpet}Analytical trumpet slices}
\begin{figure}[t]
    \centering
    \includegraphics[width=\linewidth]{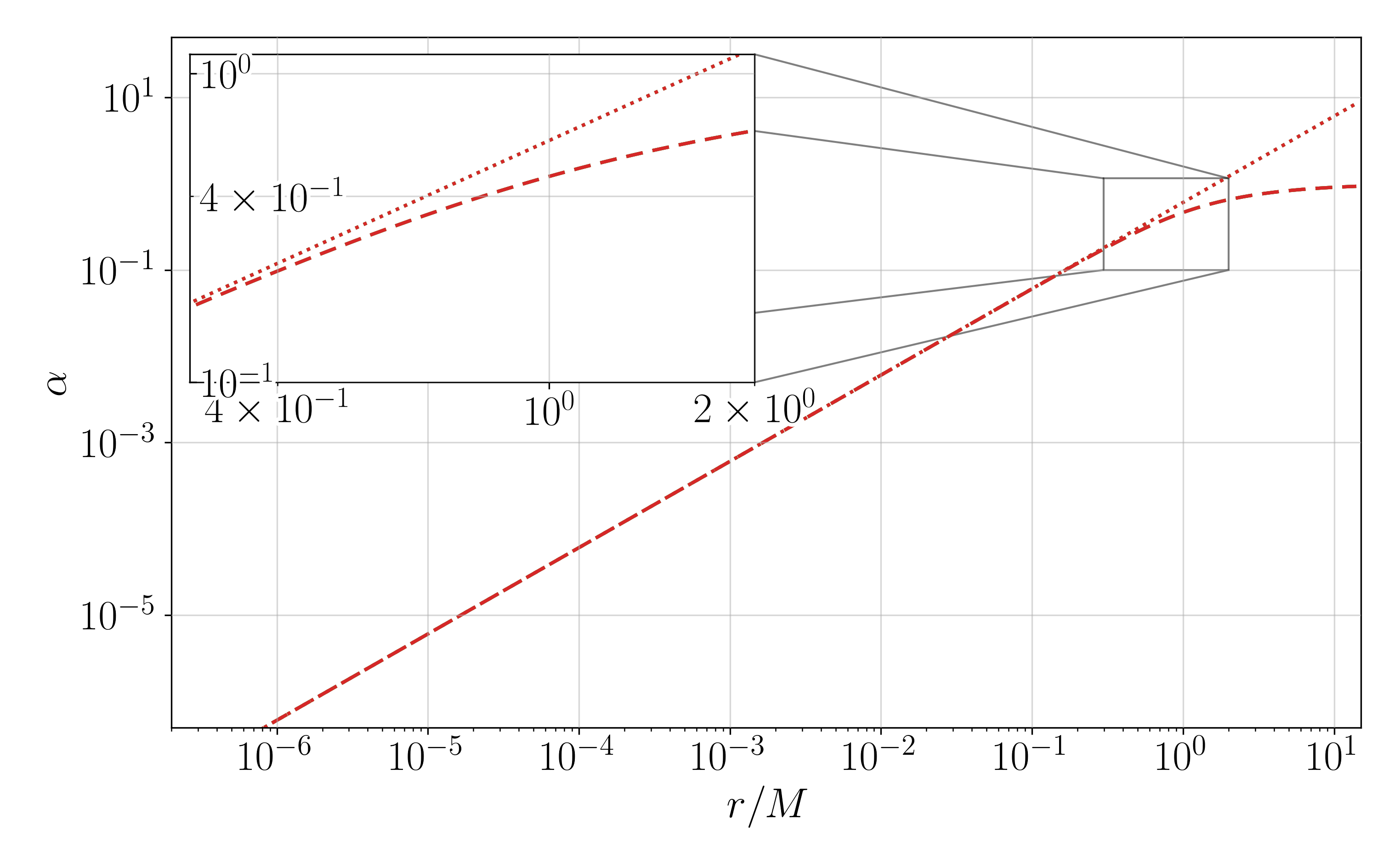}
    \caption{The analytical profile of the lapse $\alpha(r)$ for analytical trumpet slices does not depend on the black-hole charge.}
    \label{fig:trumpet-lapse}
\end{figure}

We next consider
\begin{eqnarray} \label{eq:f_analytical}
    f(\alpha) = \dfrac{1 - \alpha}{\alpha},
\end{eqnarray}
which, for uncharged black holes, results in the completely analytical trumpet slices of \cite{DennisonBaumgarte2014}.  Inserting (\ref{eq:f_analytical}) into \eqs~(\ref{eq:zero}) yields the critical lapse
\begin{eqnarray}
    \alpha_c = \dfrac{1 - \lambda^2}{2 - \lambda^2}, \label{eq:trumpet-critical-lapse}
\end{eqnarray}
together with the critical radius
\begin{eqnarray}
    \hat{R}_c = 2 - \lambda^2. \label{eq:trumpet-critical-radius}
\end{eqnarray}
With $I(\alpha) = -\ln(1 - \alpha)$, (\ref{eq:first-integral}) becomes 
\begin{eqnarray}
    \alpha^2 = 1 - \dfrac{2}{\hat{R}} + \dfrac{\lambda^2}{\hat{R}^2} + \dfrac{\hat{C}}{\hat{R}^4(1 - \alpha)^2}. \label{eq:trumpet-first-integral}
\end{eqnarray}
Substituting (\ref{eq:trumpet-critical-lapse}) and (\ref{eq:trumpet-critical-radius}) into (\ref{eq:trumpet-first-integral}), we then obtain the constant of integration
\begin{eqnarray}
    \hat{C} = 1 - \lambda^2. \label{eq:trumpet-constant}
\end{eqnarray}
Inserting the above into (\ref{eq:trumpet-first-integral}) and searching for roots of the lapse yields a quartic equation for $\hat R_0$ with two real solutions, one of which is $\hat{R}_0 = 1$. We compute
\begin{eqnarray}
    \hat{a}_1 = \dfrac{- \hat{R}_0^3 + \lambda^2\hat{R}_0^2 + 2\hat{C}}{\hat{C}\hat{R}_0} \label{eq:trumpet-lapse-derivative}
\end{eqnarray}
by implicit differentiation of (\ref{eq:trumpet-first-integral}) and find that $\hat{R}_0 = 1$ is the only root for which $\hat{a}_1 > 0$. In particular, for $\hat{R}_0 = 1$, we find $\hat{a}_1 = 1$, and hence $1/\gamma = 1$, independently of the charge-to-mass ratio $\lambda$.

Substituting (\ref{eq:trumpet-constant}) into (\ref{eq:trumpet-first-integral}), we find two real solutions for the lapse as a function of the areal radius. Only the solution
\begin{eqnarray}
    \alpha(R) = \dfrac{\hat{R} - 1}{\hat{R}}, \label{eq:trumpet-lapse}
\end{eqnarray}
however, which is identical to the extremal solution (\ref{eq:alpha_of_R_extremal}) but, remarkably, holds for all values of $\lambda$, satisfies $\alpha' > 0$ for all $\hat{R}$.  We confirm that (\ref{eq:trumpet-lapse}) agrees with the trumpet slices derived in  \cite{DennisonBaumgarteMontero2014} in the appropriate limit.

We take the solution for the lapse (\ref{eq:trumpet-lapse}) and convert from areal radius $R$ to isotropic radius $r$ as explained in Sec.~\ref{subsec:isotropic}, and show our results for the lapse $\alpha(r)$ in Fig.~\ref{fig:trumpet-lapse}.

\subsection{\label{subsec:noshock}Slices that avoid gauge shocks}

\subsubsection{\label{subsubsec:full}Full gauge-shock avoidance}
\begin{figure}[t]
    \centering
    \includegraphics[width=\linewidth]{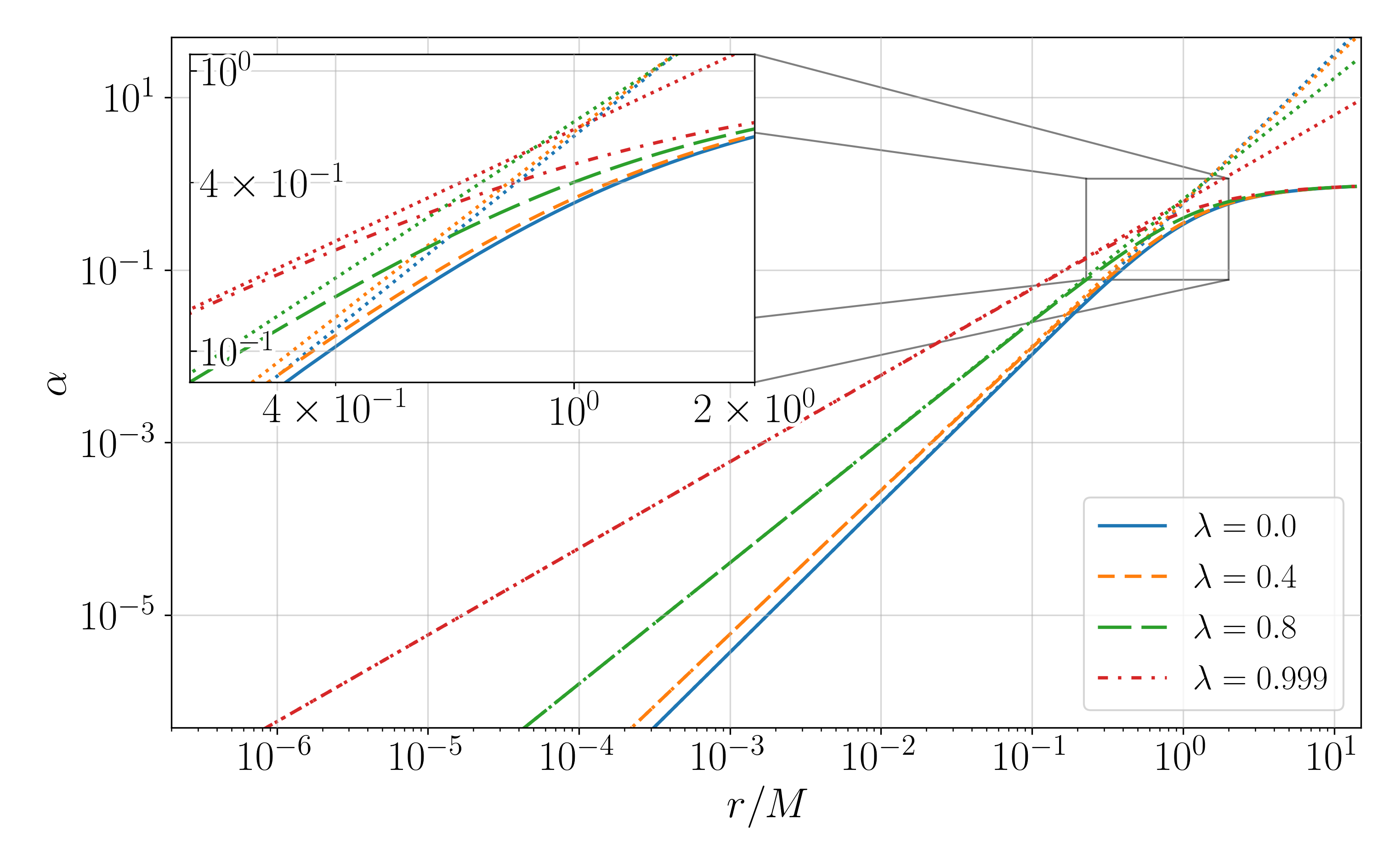}
    \caption{The analytical profile of the lapse $\alpha(r)$ for $\lambda \in \qty{0, 0.4, 0.8, 0.999}$ for fully gauge-shock-avoiding slices with $\kappa = 1$.}
    \label{fig:fully-shock-avoiding-lapse}
\end{figure}

We next consider Bona-Mass\'o functions of the form 
\begin{eqnarray}
    f(\alpha) = 1 + \dfrac{\kappa}{\alpha^2} \label{eq:fully-shock-avoiding-slices}
\end{eqnarray}
with $\kappa > 0$, which Alcubierre \cite{Alcubierre1997} proposed as an alternative to 1+log slicing that helps avoid ``gauge-shocks", i.e.~coordinate discontinuities that arise during evolution (see also \cite{JimVA21} for applications in simulations of critical collapse, and \cite{BaumgarteHilditch2022} for tests and calibrations).  

With $f(\alpha)$ given by (\ref{eq:fully-shock-avoiding-slices}), \eq~(\ref{eq:lapse-derivative}) becomes
\begin{eqnarray}
    M\alpha' = \frac{\alpha^2 + \kappa}{\alpha \hat{R}} \frac{2 - 3/\hat{R} + \lambda^2/\hat{R}^2 - 2\alpha^2}{1 - 2/\hat{R} + \lambda^2/\hat{R}^2 + \kappa}. \label{eq:fully-shock-avoiding-lapse-derivative}
\end{eqnarray}
As we had discussed in Section \ref{subsec:regularity}, the denominator of the right-hand side now vanishes for $\alpha = 0$.  For $\alpha = 0$, the numerator of (\ref{eq:fully-shock-avoiding-lapse-derivative}), i.e.~\eq~(\ref{eq:numerator-zero}), has a root for
\begin{eqnarray}
    \hat{R}_c = \dfrac{3 + \sqrt{9 - 8\lambda^2}}{4} \label{eq:fully-shock-avoiding-critical-radius-one}
\end{eqnarray}
(where we have chosen the ``outermost" solution to a quadratic equation for $\hat R_c$). 

It is possible, of course, that the denominator of the second factor in (\ref{eq:fully-shock-avoiding-lapse-derivative}) has a root for a radius larger than $\hat R_c$ as determined in (\ref{eq:fully-shock-avoiding-critical-radius-one}).  This root occurs at a radius
\begin{eqnarray}
    \hat{R}_c^{\rm alt} = \dfrac{1 + \sqrt{1 - \lambda^2\left(1 + \kappa\right)}}{1 + \kappa} \label{eq:fully-shock-avoiding-critical-radius-two}
\end{eqnarray}
(which we note exists only for $\kappa \leq 1/\lambda^2 - 1$). Substituting (\ref{eq:fully-shock-avoiding-critical-radius-two}) into (\ref{eq:numerator-zero}) we find the corresponding critical lapse
\begin{equation}
    \alpha_c^{\rm alt} = \qty(\dfrac{-1 + \sqrt{1 - \lambda^2(\kappa + 1)} - \lambda^2(\kappa - 1)}{2})^{\!\!1/2}\dfrac{1}{\lambda}. \label{eq:fully-shock-avoiding-critical-lapse-alternative}
\end{equation}
We observe that for 
\begin{eqnarray}
    \kappa > 1 - \dfrac{3 - \sqrt{9 - 8\lambda^2}}{2\lambda^2} \label{eq:second-kappa-condition}
\end{eqnarray}
no real solutions for $\alpha_c^{\rm alt}$ exist, and conclude that, in this case, the critical radius $\hat R_c$ is given by (\ref{eq:fully-shock-avoiding-critical-radius-one}) with $\alpha_c = 0$.  In the limit $\lambda \to 0$ condition (\ref{eq:second-kappa-condition}) reduces to $\kappa > 1/3$, in agreement with \cite{BaumgarteOliveira2022}. 

From here we assume that condition (\ref{eq:second-kappa-condition}) holds, and hence adopt the value (\ref{eq:fully-shock-avoiding-critical-radius-one}) for $\hat R_c$ together with $\alpha_c = 0$.  As in \cite{BaumgarteOliveira2022} we may integrate (\ref{eq:integral-of-alpha}) to obtain
\begin{eqnarray}
    I(\alpha) = \dfrac{1}{2}\ln{\qty(\frac{\alpha^2 + \kappa}{\kappa})},
\end{eqnarray}
so that (\ref{eq:first-integral}) becomes
\begin{equation} \label{eq:fully-shock-avoiding-first-integral}
   \alpha^2 = 1 - \dfrac{2}{\hat{R}} + \dfrac{\lambda^2}{\hat{R}^2} + \dfrac{\alpha^2 + \kappa}{\kappa}\dfrac{\hat{C}}{\hat{R}^4}.
\end{equation}\newline
Solving for the constant of integration $\hat C$ we obtain
\begin{eqnarray}
    \hat{C} = -\hat{R}_c^4\left(1 - \dfrac{2}{\hat{R}_c} + \frac{\lambda^2}{\hat{R}_c^2}\right), \label{eq:fully-shock-avoiding-constant-R}
\end{eqnarray}
and substituting (\ref{eq:fully-shock-avoiding-critical-radius-one}) for $\hat{R}_c$ yields 
\begin{eqnarray}
    \hat{C} = \frac{27}{32} + \frac{1}{32}\left(9 - 8\lambda^2\right)^{3/2} - \frac{9\lambda^2}{8} + \frac{\lambda^4}{4}. \label{eq:fully-shock-avoiding-constant}
\end{eqnarray}
For $\lambda = 0$, we recover $\hat{C} = 3^3/2^4$ as found by \cite{BaumgarteOliveira2022}. 

To evaluate $\hat{a}_1$, we apply L'H\^opital's rule to (\ref{eq:fully-shock-avoiding-lapse-derivative}) and impose $M\alpha' \to \hat{a}_1$ as $\alpha \to 0$ to find
\begin{eqnarray}
    \hat{a}_1 = \sqrt{\dfrac{\kappa\left(4\hat{R}_0 - 3\right)}{(1 + \kappa)\hat{R}_0^3 - 2\hat{R}_0^2 + \lambda^2\hat{R}_0}}.
\end{eqnarray}\\
Using (\ref{eq:fully-shock-avoiding-critical-radius-one}) again we then have
\begin{equation}
    \dfrac{1}{\gamma} = {\left(\frac{18\kappa + 6\kappa\sqrt{9 - 8\lambda^2} - 16\lambda^2\kappa}{ 9\kappa + (3\kappa - 1)\sqrt{9 - 8\lambda^2} + 4\lambda^2(1 - \kappa) - 3}\right)}^{\!\!1/2} \label{eq:fully-shock-avoiding-exponent}
\end{equation}
and verify that we recover 
\begin{eqnarray}
    \dfrac{1}{\gamma} = \sqrt{\dfrac{6\kappa}{3\kappa - 1}}  \label{eq:fully-shock-avoiding-exponent-no-charge}
\end{eqnarray}
for $\lambda = 0$ as in \cite{BaumgarteOliveira2022}. In the extremal limit $\lambda = 1$, (\ref{eq:fully-shock-avoiding-exponent}) reduces to $1/\gamma = 1$ independently of $\kappa$, as expected from our discussion in Section \ref{sec:extremal}.

We obtain an analytical expression for the lapse as a function of areal radius by inserting (\ref{eq:fully-shock-avoiding-constant}) into (\ref{eq:fully-shock-avoiding-first-integral}),
\begin{widetext}
\begin{eqnarray}
    \alpha(R) = \qty(\dfrac{32\hat{R}^4 - 64\hat{R}^3 + 32\lambda^2 \hat{R}^2 + 27 + \qty(9 - 8\lambda^2)^{3/2} - 36\lambda^2 + 8\lambda^4}{32\kappa \hat{R}^4 - 27 - \qty(9 - 8\lambda^2)^{3/2} + 36\lambda^2 - 8\lambda^4})^{\!\!1/2}\sqrt{\kappa}.
\end{eqnarray}
\end{widetext}

We use the above solution to transform from areal radius $R$ to isotropic radius $r$ as described in Section \ref{subsec:isotropic}, and show results for the lapse $\alpha(r)$ in Fig.~\ref{fig:fully-shock-avoiding-lapse}.

\subsubsection{\label{subsubsec:leading}Shock-avoidance to leading order}
\begin{figure}[t]
    \centering
    \includegraphics[width=\linewidth]{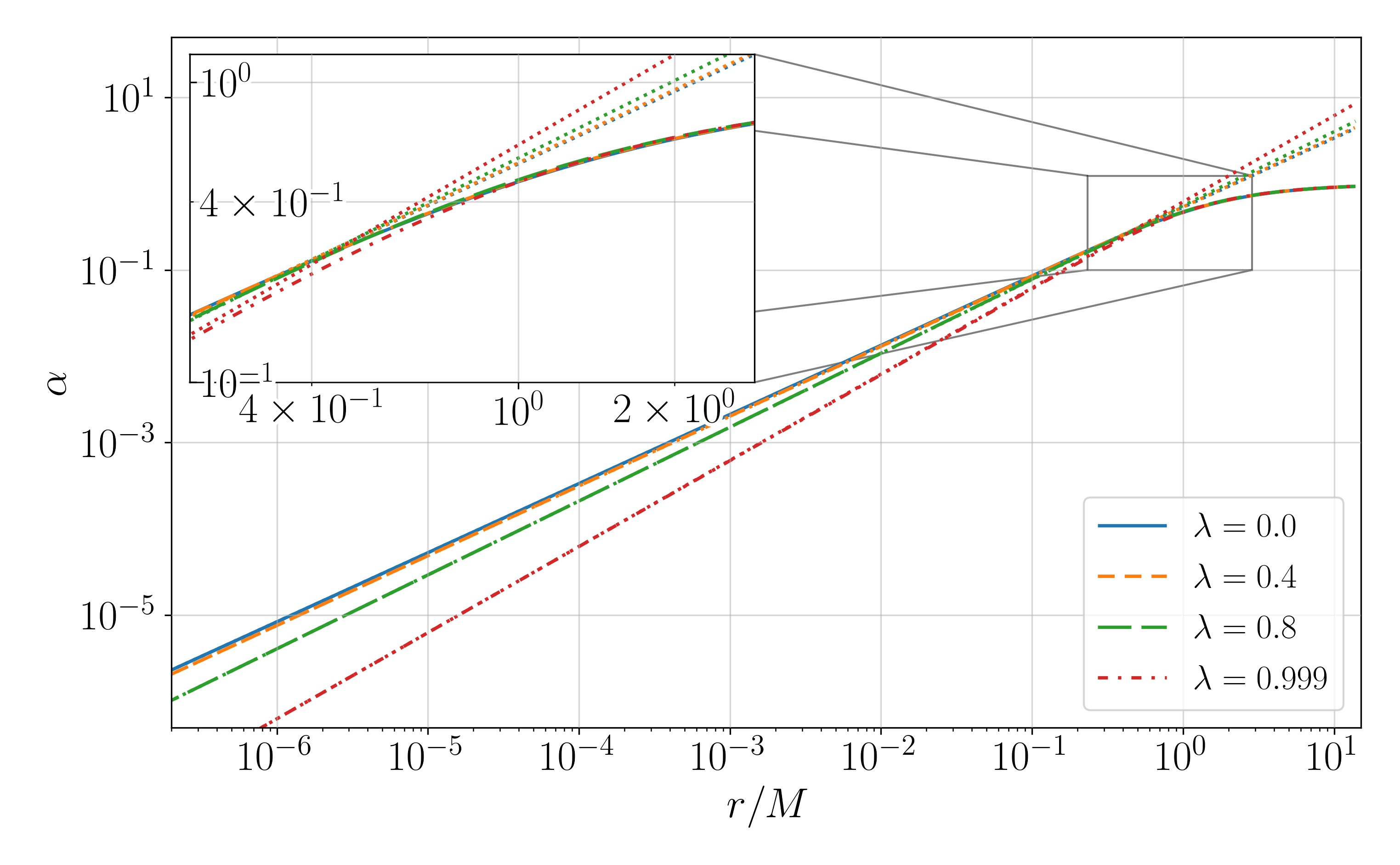}
    \caption{The numerical profile of the lapse $\alpha(r)$ for $\lambda \in \qty{0, 0.4, 0.8, 0.999}$ for zeroth-order shock-avoiding slices with $a_0 = 4/3$.}
    \label{fig:shock-avoiding-lapse}
\end{figure}

The fully shock-avoiding slicing condition given by (\ref{eq:fully-shock-avoiding-slices}) has the unusual property that it allows the lapse function to become negative during a numerical evolution (see \cite{Alcubierre1997,Alc03,BaumgarteHilditch2022}).  Following \cite{Alc03} we therefore consider a ``leading-order" shock-avoiding condition
\begin{eqnarray}
    f(\alpha) = \dfrac{a_0^2}{2\alpha + \left(a_0 - 2\right)\alpha^2} \label{eq:leading-order-shock-avoiding-slices}
\end{eqnarray}
(see, e.g., \cite{HeaRLZ16,RucHLZ17} for numerical applications).  We note that 1+log slicing (\ref{eq:onepluslog}) with $k = 2$ is a member of this family for $a_0 = 2$.

As in \cite{BaumgarteOliveira2022} we can find the integral (\ref{eq:first-integral}) analytically,
\begin{eqnarray}
    I(\alpha) = \dfrac{\alpha}{2a_0^2}\left(4 - (a_0 - 2)\alpha\right), \label{eq:leading-order-shock-avoiding-integral}
\end{eqnarray}
and may therefore evaluate the derivative of the lapse at its root to find
\begin{eqnarray}
    \hat{a}_1 = a_0^2\dfrac{-2\hat{R}_0^3 + 2\lambda^2\hat{R}_0^2 + 4\hat{C}}{4\hat{C}\hat{R}_0}. \label{eq:leading-order-shock-avoiding-derivative}
\end{eqnarray}
Expressions for the critical point $(\alpha_c, \hat{R}_c)$, the constant $\hat{C}$, and the root of the lapse $\hat{R}_0$, however, are more complicated, and we therefore find these quantities numerically (see Figs.~\ref{fig:all-alpha_c-C-gamma} and \ref{fig:all-R_c-R_0}). 

Using these values, we compute the lapse by applying numerical root-finding to (\ref{eq:first-integral}), then transform from areal to isotropic radius as in Sec.~\ref{subsec:isotropic}, and plot the lapse profile $\alpha(r(R))$ for a few chosen $\lambda$ in Fig. \ref{fig:shock-avoiding-lapse}.

\section{\label{sec:summary}Summary}

The 1+log slicing condition has been extremely successful in many numerical relativity simulations, including simulations of black holes and their binaries.  Our understanding and interpretation of these simulations have greatly benefited from analytical studies that applied this and other slicing conditions to single, static, and spherically symmetric black holes, i.e.~the Schwarzschild spacetime (e.g., \cite{HanHBGSO07,HanHOBO08,BauN07,Brugmann2009,BaumgarteOliveira2022}).  

Motivated by recent simulations of charged black holes and their interactions we generalize some of the above treatments by applying them to the charged counterpart of Schwarzschild black holes, namely Reissner-Nordstr\"om spacetimes. In addition to 1+log slicing we consider several other slicing conditions, specified by their corresponding Bona-Mass\'o functions $f(\alpha)$, that have been adopted in numerical simulations. For some of these conditions the slices can be constructed analytically, while for others we use numerical root-finding to solve rather unwieldy quartic equations. We identify critical parameters for these slices, parameterized by the charge-to-mass ratio $\lambda = Q/M$, and transform to isotropic coordinates as they would likely be adopted in numerical simulations.  In particular we observe that, in the extremal limit $\lambda \to 1$, all slices approach a unique slice that is independent of the Bona-Mass\'o functions considered in this paper,  as we anticipate in Sec.~\ref{sec:extremal}.

\begin{acknowledgements}
SEL acknowledges support through an undergraduate research fellowship at Bowdoin College, and HPO would like to thank Bowdoin College and its Department of Physics and Astronomy for hospitality. This work was supported in part by National Science Foundation (NSF) grant PHY-2010394 to Bowdoin College and the Coordena\c c\~ao de Aperfei\c coamento de Pessoal de N\'ivel Superior - Brasil (CAPES) - Finance Code 001.
\end{acknowledgements}

\bibliography{references}

\providecommand{\noopsort}[1]{}\providecommand{\singleletter}[1]{#1}%
\begin{thebibliography}{35}%
\makeatletter
\providecommand \@ifxundefined [1]{%
 \@ifx{#1\undefined}
}%
\providecommand \@ifnum [1]{%
 \ifnum #1\expandafter \@firstoftwo
 \else \expandafter \@secondoftwo
 \fi
}%
\providecommand \@ifx [1]{%
 \ifx #1\expandafter \@firstoftwo
 \else \expandafter \@secondoftwo
 \fi
}%
\providecommand \natexlab [1]{#1}%
\providecommand \enquote  [1]{``#1''}%
\providecommand \bibnamefont  [1]{#1}%
\providecommand \bibfnamefont [1]{#1}%
\providecommand \citenamefont [1]{#1}%
\providecommand \href@noop [0]{\@secondoftwo}%
\providecommand \href [0]{\begingroup \@sanitize@url \@href}%
\providecommand \@href[1]{\@@startlink{#1}\@@href}%
\providecommand \@@href[1]{\endgroup#1\@@endlink}%
\providecommand \@sanitize@url [0]{\catcode `\\12\catcode `\$12\catcode
  `\&12\catcode `\#12\catcode `\^12\catcode `\_12\catcode `\%12\relax}%
\providecommand \@@startlink[1]{}%
\providecommand \@@endlink[0]{}%
\providecommand \url  [0]{\begingroup\@sanitize@url \@url }%
\providecommand \@url [1]{\endgroup\@href {#1}{\urlprefix }}%
\providecommand \urlprefix  [0]{URL }%
\providecommand \Eprint [0]{\href }%
\providecommand \doibase [0]{https://doi.org/}%
\providecommand \selectlanguage [0]{\@gobble}%
\providecommand \bibinfo  [0]{\@secondoftwo}%
\providecommand \bibfield  [0]{\@secondoftwo}%
\providecommand \translation [1]{[#1]}%
\providecommand \BibitemOpen [0]{}%
\providecommand \bibitemStop [0]{}%
\providecommand \bibitemNoStop [0]{.\EOS\space}%
\providecommand \EOS [0]{\spacefactor3000\relax}%
\providecommand \BibitemShut  [1]{\csname bibitem#1\endcsname}%
\let\auto@bib@innerbib\@empty
\bibitem [{\citenamefont {Bona}\ \emph {et~al.}(1995)\citenamefont {Bona},
  \citenamefont {Mass\'{o}}, \citenamefont {Seidel},\ and\ \citenamefont
  {Stela}}]{BonaMasso1994}%
  \BibitemOpen
  \bibfield  {author} {\bibinfo {author} {\bibfnamefont {C.}~\bibnamefont
  {Bona}}, \bibinfo {author} {\bibfnamefont {J.}~\bibnamefont {Mass\'{o}}},
  \bibinfo {author} {\bibfnamefont {E.}~\bibnamefont {Seidel}},\ and\ \bibinfo
  {author} {\bibfnamefont {J.}~\bibnamefont {Stela}},\ }\bibfield  {title}
  {\bibinfo {title} {{A new formalism for numerical relativity}},\ }\href
  {https://doi.org/10.1103/PhysRevLett.75.600} {\bibfield  {journal} {\bibinfo
  {journal} {Phys. Rev. Lett.}\ }\textbf {\bibinfo {volume} {75}},\ \bibinfo
  {pages} {600} (\bibinfo {year} {1995})}\BibitemShut {NoStop}%
\bibitem [{\citenamefont {{Alcubierre}}\ and\ \citenamefont
  {{Br{\"u}gmann}}(2001)}]{AlcB01}%
  \BibitemOpen
  \bibfield  {author} {\bibinfo {author} {\bibfnamefont {M.}~\bibnamefont
  {{Alcubierre}}}\ and\ \bibinfo {author} {\bibfnamefont {B.}~\bibnamefont
  {{Br{\"u}gmann}}},\ }\bibfield  {title} {\bibinfo {title} {{Simple excision
  of a black hole in 3+1 numerical relativity}},\ }\href
  {https://doi.org/10.1103/PhysRevD.63.104006} {\bibfield  {journal} {\bibinfo
  {journal} {\prd}\ }\textbf {\bibinfo {volume} {63}},\ \bibinfo {eid} {104006}
  (\bibinfo {year} {2001})},\ \Eprint {https://arxiv.org/abs/gr-qc/0008067}
  {arXiv:gr-qc/0008067 [gr-qc]} \BibitemShut {NoStop}%
\bibitem [{\citenamefont {{Alcubierre}}\ \emph {et~al.}(2003)\citenamefont
  {{Alcubierre}}, \citenamefont {{Br{\"u}gmann}}, \citenamefont {{Diener}},
  \citenamefont {{Koppitz}}, \citenamefont {{Pollney}}, \citenamefont
  {{Seidel}},\ and\ \citenamefont {{Takahashi}}}]{AlcBDKPST03}%
  \BibitemOpen
  \bibfield  {author} {\bibinfo {author} {\bibfnamefont {M.}~\bibnamefont
  {{Alcubierre}}}, \bibinfo {author} {\bibfnamefont {B.}~\bibnamefont
  {{Br{\"u}gmann}}}, \bibinfo {author} {\bibfnamefont {P.}~\bibnamefont
  {{Diener}}}, \bibinfo {author} {\bibfnamefont {M.}~\bibnamefont {{Koppitz}}},
  \bibinfo {author} {\bibfnamefont {D.}~\bibnamefont {{Pollney}}}, \bibinfo
  {author} {\bibfnamefont {E.}~\bibnamefont {{Seidel}}},\ and\ \bibinfo
  {author} {\bibfnamefont {R.}~\bibnamefont {{Takahashi}}},\ }\bibfield
  {title} {\bibinfo {title} {{Gauge conditions for long-term numerical black
  hole evolutions without excision}},\ }\href
  {https://doi.org/10.1103/PhysRevD.67.084023} {\bibfield  {journal} {\bibinfo
  {journal} {\prd}\ }\textbf {\bibinfo {volume} {67}},\ \bibinfo {eid} {084023}
  (\bibinfo {year} {2003})},\ \Eprint {https://arxiv.org/abs/gr-qc/0206072}
  {arXiv:gr-qc/0206072 [gr-qc]} \BibitemShut {NoStop}%
\bibitem [{\citenamefont {{van Meter}}\ \emph {et~al.}(2006)\citenamefont {{van
  Meter}}, \citenamefont {{Baker}}, \citenamefont {{Koppitz}},\ and\
  \citenamefont {{Choi}}}]{vanMBKC06}%
  \BibitemOpen
  \bibfield  {author} {\bibinfo {author} {\bibfnamefont {J.~R.}\ \bibnamefont
  {{van Meter}}}, \bibinfo {author} {\bibfnamefont {J.~G.}\ \bibnamefont
  {{Baker}}}, \bibinfo {author} {\bibfnamefont {M.}~\bibnamefont {{Koppitz}}},\
  and\ \bibinfo {author} {\bibfnamefont {D.-I.}\ \bibnamefont {{Choi}}},\
  }\bibfield  {title} {\bibinfo {title} {{How to move a black hole without
  excision: Gauge conditions for the numerical evolution of a moving
  puncture}},\ }\href {https://doi.org/10.1103/PhysRevD.73.124011} {\bibfield
  {journal} {\bibinfo  {journal} {\prd}\ }\textbf {\bibinfo {volume} {73}},\
  \bibinfo {eid} {124011} (\bibinfo {year} {2006})},\ \Eprint
  {https://arxiv.org/abs/gr-qc/0605030} {arXiv:gr-qc/0605030 [gr-qc]}
  \BibitemShut {NoStop}%
\bibitem [{\citenamefont {{Campanelli}}\ \emph {et~al.}(2006)\citenamefont
  {{Campanelli}}, \citenamefont {{Lousto}}, \citenamefont {{Marronetti}},\ and\
  \citenamefont {{Zlochower}}}]{CamLMZ06}%
  \BibitemOpen
  \bibfield  {author} {\bibinfo {author} {\bibfnamefont {M.}~\bibnamefont
  {{Campanelli}}}, \bibinfo {author} {\bibfnamefont {C.~O.}\ \bibnamefont
  {{Lousto}}}, \bibinfo {author} {\bibfnamefont {P.}~\bibnamefont
  {{Marronetti}}},\ and\ \bibinfo {author} {\bibfnamefont {Y.}~\bibnamefont
  {{Zlochower}}},\ }\bibfield  {title} {\bibinfo {title} {{Accurate Evolutions
  of Orbiting Black-Hole Binaries without Excision}},\ }\href
  {https://doi.org/10.1103/PhysRevLett.96.111101} {\bibfield  {journal}
  {\bibinfo  {journal} {\prl}\ }\textbf {\bibinfo {volume} {96}},\ \bibinfo
  {eid} {111101} (\bibinfo {year} {2006})},\ \Eprint
  {https://arxiv.org/abs/gr-qc/0511048} {arXiv:gr-qc/0511048 [gr-qc]}
  \BibitemShut {NoStop}%
\bibitem [{\citenamefont {{Baker}}\ \emph {et~al.}(2006)\citenamefont
  {{Baker}}, \citenamefont {{Centrella}}, \citenamefont {{Choi}}, \citenamefont
  {{Koppitz}},\ and\ \citenamefont {{van Meter}}}]{BakCCKM06}%
  \BibitemOpen
  \bibfield  {author} {\bibinfo {author} {\bibfnamefont {J.~G.}\ \bibnamefont
  {{Baker}}}, \bibinfo {author} {\bibfnamefont {J.}~\bibnamefont
  {{Centrella}}}, \bibinfo {author} {\bibfnamefont {D.-I.}\ \bibnamefont
  {{Choi}}}, \bibinfo {author} {\bibfnamefont {M.}~\bibnamefont {{Koppitz}}},\
  and\ \bibinfo {author} {\bibfnamefont {J.}~\bibnamefont {{van Meter}}},\
  }\bibfield  {title} {\bibinfo {title} {{Gravitational-Wave Extraction from an
  Inspiraling Configuration of Merging Black Holes}},\ }\href
  {https://doi.org/10.1103/PhysRevLett.96.111102} {\bibfield  {journal}
  {\bibinfo  {journal} {\prl}\ }\textbf {\bibinfo {volume} {96}},\ \bibinfo
  {eid} {111102} (\bibinfo {year} {2006})},\ \Eprint
  {https://arxiv.org/abs/gr-qc/0511103} {arXiv:gr-qc/0511103 [gr-qc]}
  \BibitemShut {NoStop}%
\bibitem [{\citenamefont {{Hannam}}\ \emph
  {et~al.}(2007{\natexlab{a}})\citenamefont {{Hannam}}, \citenamefont {{Husa}},
  \citenamefont {{Pollney}}, \citenamefont {{Br{\"u}gmann}},\ and\
  \citenamefont {{Murchadha}}}]{HanHPBM07}%
  \BibitemOpen
  \bibfield  {author} {\bibinfo {author} {\bibfnamefont {M.}~\bibnamefont
  {{Hannam}}}, \bibinfo {author} {\bibfnamefont {S.}~\bibnamefont {{Husa}}},
  \bibinfo {author} {\bibfnamefont {D.}~\bibnamefont {{Pollney}}}, \bibinfo
  {author} {\bibfnamefont {B.}~\bibnamefont {{Br{\"u}gmann}}},\ and\ \bibinfo
  {author} {\bibfnamefont {N.~{\'O}.}\ \bibnamefont {{Murchadha}}},\ }\bibfield
   {title} {\bibinfo {title} {{Geometry and Regularity of Moving Punctures}},\
  }\href {https://doi.org/10.1103/PhysRevLett.99.241102} {\bibfield  {journal}
  {\bibinfo  {journal} {\prl}\ }\textbf {\bibinfo {volume} {99}},\ \bibinfo
  {eid} {241102} (\bibinfo {year} {2007}{\natexlab{a}})},\ \Eprint
  {https://arxiv.org/abs/gr-qc/0606099} {arXiv:gr-qc/0606099 [gr-qc]}
  \BibitemShut {NoStop}%
\bibitem [{\citenamefont {{Hannam}}\ \emph
  {et~al.}(2007{\natexlab{b}})\citenamefont {{Hannam}}, \citenamefont {{Husa}},
  \citenamefont {{Br{\"u}gmann}}, \citenamefont {{Gonz{\'a}lez}}, \citenamefont
  {{Sperhake}},\ and\ \citenamefont {{Murchadha}}}]{HanHBGSO07}%
  \BibitemOpen
  \bibfield  {author} {\bibinfo {author} {\bibfnamefont {M.}~\bibnamefont
  {{Hannam}}}, \bibinfo {author} {\bibfnamefont {S.}~\bibnamefont {{Husa}}},
  \bibinfo {author} {\bibfnamefont {B.}~\bibnamefont {{Br{\"u}gmann}}},
  \bibinfo {author} {\bibfnamefont {J.~A.}\ \bibnamefont {{Gonz{\'a}lez}}},
  \bibinfo {author} {\bibfnamefont {U.}~\bibnamefont {{Sperhake}}},\ and\
  \bibinfo {author} {\bibfnamefont {N.~{\'O}.}\ \bibnamefont {{Murchadha}}},\
  }\bibfield  {title} {\bibinfo {title} {{Where do moving punctures go?}},\
  }in\ \href {https://doi.org/10.1088/1742-6596/66/1/012047} {\emph {\bibinfo
  {booktitle} {Journal of Physics Conference Series}}},\ \bibinfo {series}
  {Journal of Physics Conference Series}, Vol.~\bibinfo {volume} {66}\
  (\bibinfo {year} {2007})\ p.\ \bibinfo {pages} {012047},\ \Eprint
  {https://arxiv.org/abs/gr-qc/0612097} {arXiv:gr-qc/0612097 [gr-qc]}
  \BibitemShut {NoStop}%
\bibitem [{\citenamefont {{Baumgarte}}\ and\ \citenamefont
  {{Naculich}}(2007)}]{BauN07}%
  \BibitemOpen
  \bibfield  {author} {\bibinfo {author} {\bibfnamefont {T.~W.}\ \bibnamefont
  {{Baumgarte}}}\ and\ \bibinfo {author} {\bibfnamefont {S.~G.}\ \bibnamefont
  {{Naculich}}},\ }\bibfield  {title} {\bibinfo {title} {{Analytical
  representation of a black hole puncture solution}},\ }\href
  {https://doi.org/10.1103/PhysRevD.75.067502} {\bibfield  {journal} {\bibinfo
  {journal} {\prd}\ }\textbf {\bibinfo {volume} {75}},\ \bibinfo {eid} {067502}
  (\bibinfo {year} {2007})},\ \Eprint {https://arxiv.org/abs/gr-qc/0701037}
  {arXiv:gr-qc/0701037 [gr-qc]} \BibitemShut {NoStop}%
\bibitem [{\citenamefont {{Hannam}}\ \emph {et~al.}(2008)\citenamefont
  {{Hannam}}, \citenamefont {{Husa}}, \citenamefont {{Ohme}}, \citenamefont
  {{Br{\"u}gmann}},\ and\ \citenamefont {{{\'O} Murchadha}}}]{HanHOBO08}%
  \BibitemOpen
  \bibfield  {author} {\bibinfo {author} {\bibfnamefont {M.}~\bibnamefont
  {{Hannam}}}, \bibinfo {author} {\bibfnamefont {S.}~\bibnamefont {{Husa}}},
  \bibinfo {author} {\bibfnamefont {F.}~\bibnamefont {{Ohme}}}, \bibinfo
  {author} {\bibfnamefont {B.}~\bibnamefont {{Br{\"u}gmann}}},\ and\ \bibinfo
  {author} {\bibfnamefont {N.}~\bibnamefont {{{\'O} Murchadha}}},\ }\bibfield
  {title} {\bibinfo {title} {{Wormholes and trumpets: Schwarzschild spacetime
  for the moving-puncture generation}},\ }\href
  {https://doi.org/10.1103/PhysRevD.78.064020} {\bibfield  {journal} {\bibinfo
  {journal} {\prd}\ }\textbf {\bibinfo {volume} {78}},\ \bibinfo {eid} {064020}
  (\bibinfo {year} {2008})},\ \Eprint {https://arxiv.org/abs/0804.0628}
  {arXiv:0804.0628 [gr-qc]} \BibitemShut {NoStop}%
\bibitem [{\citenamefont {Br{\"u}gmann}(2009)}]{Brugmann2009}%
  \BibitemOpen
  \bibfield  {author} {\bibinfo {author} {\bibfnamefont {B.}~\bibnamefont
  {Br{\"u}gmann}},\ }\bibfield  {title} {\bibinfo {title} {{Schwarzschild black
  hole as moving puncture in isotropic coordinates}},\ }\href
  {https://doi.org/10.1007/s10714-009-0818-6} {\bibfield  {journal} {\bibinfo
  {journal} {Gen. Rel. Grav.}\ }\textbf {\bibinfo {volume} {41}},\ \bibinfo
  {pages} {2131} (\bibinfo {year} {2009})}\BibitemShut {NoStop}%
\bibitem [{\citenamefont {Alcubierre}\ \emph {et~al.}(2009)\citenamefont
  {Alcubierre}, \citenamefont {Degollado},\ and\ \citenamefont
  {Salgado}}]{Alcubierre2009}%
  \BibitemOpen
  \bibfield  {author} {\bibinfo {author} {\bibfnamefont {M.}~\bibnamefont
  {Alcubierre}}, \bibinfo {author} {\bibfnamefont {J.~C.}\ \bibnamefont
  {Degollado}},\ and\ \bibinfo {author} {\bibfnamefont {M.}~\bibnamefont
  {Salgado}},\ }\bibfield  {title} {\bibinfo {title} {Einstein-maxwell system
  in $3+1$ form and initial data for multiple charged black holes},\ }\href
  {https://doi.org/10.1103/PhysRevD.80.104022} {\bibfield  {journal} {\bibinfo
  {journal} {Phys. Rev. D}\ }\textbf {\bibinfo {volume} {80}},\ \bibinfo
  {pages} {104022} (\bibinfo {year} {2009})}\BibitemShut {NoStop}%
\bibitem [{\citenamefont {Zilhao}\ \emph {et~al.}(2012)\citenamefont {Zilhao},
  \citenamefont {Cardoso}, \citenamefont {Herdeiro}, \citenamefont {Lehner},\
  and\ \citenamefont {Sperhake}}]{Zilhao2012}%
  \BibitemOpen
  \bibfield  {author} {\bibinfo {author} {\bibfnamefont {M.}~\bibnamefont
  {Zilhao}}, \bibinfo {author} {\bibfnamefont {V.}~\bibnamefont {Cardoso}},
  \bibinfo {author} {\bibfnamefont {C.}~\bibnamefont {Herdeiro}}, \bibinfo
  {author} {\bibfnamefont {L.}~\bibnamefont {Lehner}},\ and\ \bibinfo {author}
  {\bibfnamefont {U.}~\bibnamefont {Sperhake}},\ }\bibfield  {title} {\bibinfo
  {title} {{Collisions of charged black holes}},\ }\href
  {https://doi.org/10.1103/PhysRevD.85.124062} {\bibfield  {journal} {\bibinfo
  {journal} {Phys. Rev. D}\ }\textbf {\bibinfo {volume} {85}},\ \bibinfo
  {pages} {124062} (\bibinfo {year} {2012})},\ \Eprint
  {https://arxiv.org/abs/1205.1063} {arXiv:1205.1063 [gr-qc]} \BibitemShut
  {NoStop}%
\bibitem [{\citenamefont {Zilh\~ao}\ \emph
  {et~al.}(2014{\natexlab{a}})\citenamefont {Zilh\~ao}, \citenamefont
  {Cardoso}, \citenamefont {Herdeiro}, \citenamefont {Lehner},\ and\
  \citenamefont {Sperhake}}]{Zilhao2013}%
  \BibitemOpen
  \bibfield  {author} {\bibinfo {author} {\bibfnamefont {M.}~\bibnamefont
  {Zilh\~ao}}, \bibinfo {author} {\bibfnamefont {V.}~\bibnamefont {Cardoso}},
  \bibinfo {author} {\bibfnamefont {C.}~\bibnamefont {Herdeiro}}, \bibinfo
  {author} {\bibfnamefont {L.}~\bibnamefont {Lehner}},\ and\ \bibinfo {author}
  {\bibfnamefont {U.}~\bibnamefont {Sperhake}},\ }\bibfield  {title} {\bibinfo
  {title} {{Collisions of oppositely charged black holes}},\ }\href
  {https://doi.org/10.1103/PhysRevD.89.044008} {\bibfield  {journal} {\bibinfo
  {journal} {Phys. Rev. D}\ }\textbf {\bibinfo {volume} {89}},\ \bibinfo
  {pages} {044008} (\bibinfo {year} {2014}{\natexlab{a}})},\ \Eprint
  {https://arxiv.org/abs/1311.6483} {arXiv:1311.6483 [gr-qc]} \BibitemShut
  {NoStop}%
\bibitem [{\citenamefont {Zilh\~ao}\ \emph
  {et~al.}(2014{\natexlab{b}})\citenamefont {Zilh\~ao}, \citenamefont
  {Cardoso}, \citenamefont {Herdeiro}, \citenamefont {Lehner},\ and\
  \citenamefont {Sperhake}}]{Zilhao2014}%
  \BibitemOpen
  \bibfield  {author} {\bibinfo {author} {\bibfnamefont {M.}~\bibnamefont
  {Zilh\~ao}}, \bibinfo {author} {\bibfnamefont {V.}~\bibnamefont {Cardoso}},
  \bibinfo {author} {\bibfnamefont {C.}~\bibnamefont {Herdeiro}}, \bibinfo
  {author} {\bibfnamefont {L.}~\bibnamefont {Lehner}},\ and\ \bibinfo {author}
  {\bibfnamefont {U.}~\bibnamefont {Sperhake}},\ }\bibfield  {title} {\bibinfo
  {title} {{Head-On Collisions of Charged Black Holes from Rest}},\ }\href
  {https://doi.org/10.1007/978-3-642-40157-2_69} {\bibfield  {journal}
  {\bibinfo  {journal} {Springer Proc. Math. Stat.}\ }\textbf {\bibinfo
  {volume} {60}},\ \bibinfo {pages} {451} (\bibinfo {year}
  {2014}{\natexlab{b}})}\BibitemShut {NoStop}%
\bibitem [{\citenamefont {Zilh\~ao}\ \emph {et~al.}(2015)\citenamefont
  {Zilh\~ao}, \citenamefont {Cardoso}, \citenamefont {Herdeiro}, \citenamefont
  {Lehner},\ and\ \citenamefont {Sperhake}}]{Zilhao2015}%
  \BibitemOpen
  \bibfield  {author} {\bibinfo {author} {\bibfnamefont {M.}~\bibnamefont
  {Zilh\~ao}}, \bibinfo {author} {\bibfnamefont {V.}~\bibnamefont {Cardoso}},
  \bibinfo {author} {\bibfnamefont {C.}~\bibnamefont {Herdeiro}}, \bibinfo
  {author} {\bibfnamefont {L.}~\bibnamefont {Lehner}},\ and\ \bibinfo {author}
  {\bibfnamefont {U.}~\bibnamefont {Sperhake}},\ }\bibfield  {title} {\bibinfo
  {title} {{Dynamics of Charged Black Holes}},\ }in\ \href
  {https://doi.org/10.1142/9789814623995_0061} {\emph {\bibinfo {booktitle}
  {{13th Marcel Grossmann Meeting on Recent Developments in Theoretical and
  Experimental General Relativity, Astrophysics, and Relativistic Field
  Theories}}}}\ (\bibinfo {year} {2015})\ pp.\ \bibinfo {pages}
  {983--985}\BibitemShut {NoStop}%
\bibitem [{\citenamefont {Jai-akson}\ \emph {et~al.}(2017)\citenamefont
  {Jai-akson}, \citenamefont {Chatrabhuti}, \citenamefont {Evnin},\ and\
  \citenamefont {Lehner}}]{Lehner2017}%
  \BibitemOpen
  \bibfield  {author} {\bibinfo {author} {\bibfnamefont {P.}~\bibnamefont
  {Jai-akson}}, \bibinfo {author} {\bibfnamefont {A.}~\bibnamefont
  {Chatrabhuti}}, \bibinfo {author} {\bibfnamefont {O.}~\bibnamefont {Evnin}},\
  and\ \bibinfo {author} {\bibfnamefont {L.}~\bibnamefont {Lehner}},\
  }\bibfield  {title} {\bibinfo {title} {{Black hole merger estimates in
  Einstein-Maxwell and Einstein-Maxwell-dilaton gravity}},\ }\href
  {https://doi.org/10.1103/PhysRevD.96.044031} {\bibfield  {journal} {\bibinfo
  {journal} {Phys. Rev. D}\ }\textbf {\bibinfo {volume} {96}},\ \bibinfo
  {pages} {044031} (\bibinfo {year} {2017})},\ \Eprint
  {https://arxiv.org/abs/1706.06519} {arXiv:1706.06519 [gr-qc]} \BibitemShut
  {NoStop}%
\bibitem [{\citenamefont {Bozzola}\ and\ \citenamefont
  {Paschalidis}(2021{\natexlab{a}})}]{BozzolaPaschalidis2021}%
  \BibitemOpen
  \bibfield  {author} {\bibinfo {author} {\bibfnamefont {G.}~\bibnamefont
  {Bozzola}}\ and\ \bibinfo {author} {\bibfnamefont {V.}~\bibnamefont
  {Paschalidis}},\ }\bibfield  {title} {\bibinfo {title} {General relativistic
  simulations of the quasicircular inspiral and merger of charged black holes:
  {GW150914} and fundamental physics implications},\ }\href
  {https://doi.org/10.1103/PhysRevLett.126.041103} {\bibfield  {journal}
  {\bibinfo  {journal} {Phys. Rev. Lett.}\ }\textbf {\bibinfo {volume} {126}},\
  \bibinfo {pages} {041103} (\bibinfo {year} {2021}{\natexlab{a}})}\BibitemShut
  {NoStop}%
\bibitem [{\citenamefont {Bozzola}\ and\ \citenamefont
  {Paschalidis}(2021{\natexlab{b}})}]{Bozzola2021}%
  \BibitemOpen
  \bibfield  {author} {\bibinfo {author} {\bibfnamefont {G.}~\bibnamefont
  {Bozzola}}\ and\ \bibinfo {author} {\bibfnamefont {V.}~\bibnamefont
  {Paschalidis}},\ }\bibfield  {title} {\bibinfo {title} {{Numerical-relativity
  simulations of the quasicircular inspiral and merger of nonspinning, charged
  black holes: Methods and comparison with approximate approaches}},\ }\href
  {https://doi.org/10.1103/PhysRevD.104.044004} {\bibfield  {journal} {\bibinfo
   {journal} {Phys. Rev. D}\ }\textbf {\bibinfo {volume} {104}},\ \bibinfo
  {pages} {044004} (\bibinfo {year} {2021}{\natexlab{b}})},\ \Eprint
  {https://arxiv.org/abs/2104.06978} {arXiv:2104.06978 [gr-qc]} \BibitemShut
  {NoStop}%
\bibitem [{\citenamefont {Bozzola}(2022)}]{Bozzola2022}%
  \BibitemOpen
  \bibfield  {author} {\bibinfo {author} {\bibfnamefont {G.}~\bibnamefont
  {Bozzola}},\ }\bibfield  {title} {\bibinfo {title} {Does charge matter in
  high-energy collisions of black holes?},\ }\href
  {https://doi.org/10.1103/PhysRevLett.128.071101} {\bibfield  {journal}
  {\bibinfo  {journal} {Phys. Rev. Lett.}\ }\textbf {\bibinfo {volume} {128}},\
  \bibinfo {pages} {071101} (\bibinfo {year} {2022})}\BibitemShut {NoStop}%
\bibitem [{\citenamefont {{Mukherjee}}\ \emph {et~al.}(2022)\citenamefont
  {{Mukherjee}}, \citenamefont {{Johnson-McDaniel}}, \citenamefont {{Tichy}},\
  and\ \citenamefont {{Liebling}}}]{Mukherjee2022}%
  \BibitemOpen
  \bibfield  {author} {\bibinfo {author} {\bibfnamefont {S.}~\bibnamefont
  {{Mukherjee}}}, \bibinfo {author} {\bibfnamefont {N.~K.}\ \bibnamefont
  {{Johnson-McDaniel}}}, \bibinfo {author} {\bibfnamefont {W.}~\bibnamefont
  {{Tichy}}},\ and\ \bibinfo {author} {\bibfnamefont {S.~L.}\ \bibnamefont
  {{Liebling}}},\ }\bibfield  {title} {\bibinfo {title} {{Conformally curved
  initial data for charged, spinning black hole binaries on arbitrary
  orbits}},\ }\href@noop {} {\bibfield  {journal} {\bibinfo  {journal} {arXiv
  e-prints}\ ,\ \bibinfo {eid} {arXiv:2202.12133}} (\bibinfo {year} {2022})},\
  \Eprint {https://arxiv.org/abs/2202.12133} {arXiv:2202.12133 [gr-qc]}
  \BibitemShut {NoStop}%
\bibitem [{\citenamefont {Luna}\ \emph {et~al.}(2022)\citenamefont {Luna},
  \citenamefont {Bozzola}, \citenamefont {Cardoso}, \citenamefont
  {Paschalidis},\ and\ \citenamefont {Zilh\~ao}}]{LunaBozzola2022}%
  \BibitemOpen
  \bibfield  {author} {\bibinfo {author} {\bibfnamefont {R.}~\bibnamefont
  {Luna}}, \bibinfo {author} {\bibfnamefont {G.}~\bibnamefont {Bozzola}},
  \bibinfo {author} {\bibfnamefont {V.}~\bibnamefont {Cardoso}}, \bibinfo
  {author} {\bibfnamefont {V.}~\bibnamefont {Paschalidis}},\ and\ \bibinfo
  {author} {\bibfnamefont {M.}~\bibnamefont {Zilh\~ao}},\ }\bibfield  {title}
  {\bibinfo {title} {Kicks in charged black hole binaries},\ }\href
  {https://doi.org/10.1103/PhysRevD.106.084017} {\bibfield  {journal} {\bibinfo
   {journal} {Phys. Rev. D}\ }\textbf {\bibinfo {volume} {106}},\ \bibinfo
  {pages} {084017} (\bibinfo {year} {2022})}\BibitemShut {NoStop}%
\bibitem [{\citenamefont {Baumgarte}\ and\ \citenamefont
  {de~Oliveira}(2022)}]{BaumgarteOliveira2022}%
  \BibitemOpen
  \bibfield  {author} {\bibinfo {author} {\bibfnamefont {T.~W.}\ \bibnamefont
  {Baumgarte}}\ and\ \bibinfo {author} {\bibfnamefont {H.~P.}\ \bibnamefont
  {de~Oliveira}},\ }\bibfield  {title} {\bibinfo {title} {{{B}ona-{M}ass{\'{o}}
  slicing conditions and the lapse close to black-hole punctures}},\ }\href
  {https://doi.org/10.1103/PhysRevD.105.064045} {\bibfield  {journal} {\bibinfo
   {journal} {Phys. Rev. D}\ }\textbf {\bibinfo {volume} {105}},\ \bibinfo
  {pages} {064045} (\bibinfo {year} {2022})}\BibitemShut {NoStop}%
\bibitem [{\citenamefont {Baumgarte}\ and\ \citenamefont
  {Shapiro}(2010)}]{BaumgarteShapiro2010}%
  \BibitemOpen
  \bibfield  {author} {\bibinfo {author} {\bibfnamefont {T.~W.}\ \bibnamefont
  {Baumgarte}}\ and\ \bibinfo {author} {\bibfnamefont {S.~L.}\ \bibnamefont
  {Shapiro}},\ }\href {https://doi.org/10.1017/CBO9781139193344} {\emph
  {\bibinfo {title} {Numerical Relativity: Solving Einstein's Equations on the
  Computer}}}\ (\bibinfo  {publisher} {Cambridge University Press},\ \bibinfo
  {year} {2010})\BibitemShut {NoStop}%
\bibitem [{\citenamefont {Reimann}\ and\ \citenamefont
  {Br{\"u}gmann}(2004)}]{Reimann2004A}%
  \BibitemOpen
  \bibfield  {author} {\bibinfo {author} {\bibfnamefont {B.}~\bibnamefont
  {Reimann}}\ and\ \bibinfo {author} {\bibfnamefont {B.}~\bibnamefont
  {Br{\"u}gmann}},\ }\bibfield  {title} {\bibinfo {title} {{Maximal slicing for
  puncture evolutions of Schwarzschild and Reissner-Nordstrom black holes}},\
  }\href {https://doi.org/10.1103/PhysRevD.69.044006} {\bibfield  {journal}
  {\bibinfo  {journal} {Phys. Rev. D}\ }\textbf {\bibinfo {volume} {69}},\
  \bibinfo {pages} {044006} (\bibinfo {year} {2004})},\ \Eprint
  {https://arxiv.org/abs/gr-qc/0307036} {arXiv:gr-qc/0307036} \BibitemShut
  {NoStop}%
\bibitem [{\citenamefont {Reimann}\ and\ \citenamefont
  {Bruegmann}(2004)}]{Reimann2004B}%
  \BibitemOpen
  \bibfield  {author} {\bibinfo {author} {\bibfnamefont {B.}~\bibnamefont
  {Reimann}}\ and\ \bibinfo {author} {\bibfnamefont {B.}~\bibnamefont
  {Bruegmann}},\ }\bibfield  {title} {\bibinfo {title} {{Late time analysis for
  maximal slicing of Reissner-Nordstrom puncture evolutions}},\ }\href
  {https://doi.org/10.1103/PhysRevD.69.124009} {\bibfield  {journal} {\bibinfo
  {journal} {Phys. Rev. D}\ }\textbf {\bibinfo {volume} {69}},\ \bibinfo
  {pages} {124009} (\bibinfo {year} {2004})},\ \Eprint
  {https://arxiv.org/abs/gr-qc/0401098} {arXiv:gr-qc/0401098} \BibitemShut
  {NoStop}%
\bibitem [{\citenamefont {Panosso~Macedo}\ \emph {et~al.}(2018)\citenamefont
  {Panosso~Macedo}, \citenamefont {Jaramillo},\ and\ \citenamefont
  {Ansorg}}]{PanossoJaramillo2018}%
  \BibitemOpen
  \bibfield  {author} {\bibinfo {author} {\bibfnamefont {R.}~\bibnamefont
  {Panosso~Macedo}}, \bibinfo {author} {\bibfnamefont {J.~L.}\ \bibnamefont
  {Jaramillo}},\ and\ \bibinfo {author} {\bibfnamefont {M.}~\bibnamefont
  {Ansorg}},\ }\bibfield  {title} {\bibinfo {title} {{Hyperboloidal slicing
  approach to quasi-normal mode expansions: the Reissner-Nordstr\"om case}},\
  }\href {https://doi.org/10.1103/PhysRevD.98.124005} {\bibfield  {journal}
  {\bibinfo  {journal} {Phys. Rev. D}\ }\textbf {\bibinfo {volume} {98}},\
  \bibinfo {pages} {124005} (\bibinfo {year} {2018})},\ \Eprint
  {https://arxiv.org/abs/1809.02837} {arXiv:1809.02837 [gr-qc]} \BibitemShut
  {NoStop}%
\bibitem [{\citenamefont {Alcubierre}(1997)}]{Alcubierre1997}%
  \BibitemOpen
  \bibfield  {author} {\bibinfo {author} {\bibfnamefont {M.}~\bibnamefont
  {Alcubierre}},\ }\bibfield  {title} {\bibinfo {title} {Appearance of
  coordinate shocks in hyperbolic formalisms of general relativity},\ }\href
  {https://doi.org/10.1103/PhysRevD.55.5981} {\bibfield  {journal} {\bibinfo
  {journal} {Phys. Rev. D}\ }\textbf {\bibinfo {volume} {55}},\ \bibinfo
  {pages} {5981} (\bibinfo {year} {1997})}\BibitemShut {NoStop}%
\bibitem [{\citenamefont {Dennison}\ and\ \citenamefont
  {Baumgarte}(2014)}]{DennisonBaumgarte2014}%
  \BibitemOpen
  \bibfield  {author} {\bibinfo {author} {\bibfnamefont {K.~A.}\ \bibnamefont
  {Dennison}}\ and\ \bibinfo {author} {\bibfnamefont {T.~W.}\ \bibnamefont
  {Baumgarte}},\ }\bibfield  {title} {\bibinfo {title} {A simple family of
  analytical trumpet slices of the {Schwarzschild} spacetime},\ }\href
  {https://doi.org/10.1088/0264-9381/31/11/117001} {\bibfield  {journal}
  {\bibinfo  {journal} {Class. Quant. Grav.}\ }\textbf {\bibinfo {volume}
  {31}},\ \bibinfo {pages} {117001} (\bibinfo {year} {2014})}\BibitemShut
  {NoStop}%
\bibitem [{\citenamefont {Dennison}\ \emph {et~al.}(2014)\citenamefont
  {Dennison}, \citenamefont {Baumgarte},\ and\ \citenamefont
  {Montero}}]{DennisonBaumgarteMontero2014}%
  \BibitemOpen
  \bibfield  {author} {\bibinfo {author} {\bibfnamefont {K.~A.}\ \bibnamefont
  {Dennison}}, \bibinfo {author} {\bibfnamefont {T.~W.}\ \bibnamefont
  {Baumgarte}},\ and\ \bibinfo {author} {\bibfnamefont {P.~J.}\ \bibnamefont
  {Montero}},\ }\bibfield  {title} {\bibinfo {title} {Trumpet slices in kerr
  spacetimes},\ }\href {https://doi.org/10.1103/PhysRevLett.113.261101}
  {\bibfield  {journal} {\bibinfo  {journal} {Phys. Rev. Lett.}\ }\textbf
  {\bibinfo {volume} {113}},\ \bibinfo {pages} {261101} (\bibinfo {year}
  {2014})}\BibitemShut {NoStop}%
\bibitem [{\citenamefont {Jim\'enez-V\'azquez}\ and\ \citenamefont
  {Alcubierre}(2022)}]{JimVA21}%
  \BibitemOpen
  \bibfield  {author} {\bibinfo {author} {\bibfnamefont {E.}~\bibnamefont
  {Jim\'enez-V\'azquez}}\ and\ \bibinfo {author} {\bibfnamefont
  {M.}~\bibnamefont {Alcubierre}},\ }\bibfield  {title} {\bibinfo {title}
  {Critical gravitational collapse of a non-minimally coupled scalar field},\
  }\href {https://doi.org/10.1103/PhysRevD.105.064071} {\bibfield  {journal}
  {\bibinfo  {journal} {Phys. Rev. D}\ }\textbf {\bibinfo {volume} {105}},\
  \bibinfo {pages} {064071} (\bibinfo {year} {2022})}\BibitemShut {NoStop}%
\bibitem [{\citenamefont {Baumgarte}\ and\ \citenamefont
  {Hilditch}(2022)}]{BaumgarteHilditch2022}%
  \BibitemOpen
  \bibfield  {author} {\bibinfo {author} {\bibfnamefont {T.~W.}\ \bibnamefont
  {Baumgarte}}\ and\ \bibinfo {author} {\bibfnamefont {D.}~\bibnamefont
  {Hilditch}},\ }\bibfield  {title} {\bibinfo {title} {Shock-avoiding slicing
  conditions: Tests and calibrations},\ }\href
  {https://doi.org/10.1103/PhysRevD.106.044014} {\bibfield  {journal} {\bibinfo
   {journal} {Phys. Rev. D}\ }\textbf {\bibinfo {volume} {106}},\ \bibinfo
  {pages} {044014} (\bibinfo {year} {2022})}\BibitemShut {NoStop}%
\bibitem [{\citenamefont {{Alcubierre}}(2003)}]{Alc03}%
  \BibitemOpen
  \bibfield  {author} {\bibinfo {author} {\bibfnamefont {M.}~\bibnamefont
  {{Alcubierre}}},\ }\bibfield  {title} {\bibinfo {title} {{Hyperbolic slicings
  of spacetime: singularity avoidance and gauge shocks}},\ }\href
  {https://doi.org/10.1088/0264-9381/20/4/304} {\bibfield  {journal} {\bibinfo
  {journal} {Classical and Quantum Gravity}\ }\textbf {\bibinfo {volume}
  {20}},\ \bibinfo {pages} {607} (\bibinfo {year} {2003})},\ \Eprint
  {https://arxiv.org/abs/gr-qc/0210050} {arXiv:gr-qc/0210050 [gr-qc]}
  \BibitemShut {NoStop}%
\bibitem [{\citenamefont {Healy}\ \emph {et~al.}(2016)\citenamefont {Healy},
  \citenamefont {Ruchlin}, \citenamefont {Lousto},\ and\ \citenamefont
  {Zlochower}}]{HeaRLZ16}%
  \BibitemOpen
  \bibfield  {author} {\bibinfo {author} {\bibfnamefont {J.}~\bibnamefont
  {Healy}}, \bibinfo {author} {\bibfnamefont {I.}~\bibnamefont {Ruchlin}},
  \bibinfo {author} {\bibfnamefont {C.~O.}\ \bibnamefont {Lousto}},\ and\
  \bibinfo {author} {\bibfnamefont {Y.}~\bibnamefont {Zlochower}},\ }\bibfield
  {title} {\bibinfo {title} {{High Energy Collisions of Black Holes Numerically
  Revisited}},\ }\href {https://doi.org/10.1103/PhysRevD.94.104020} {\bibfield
  {journal} {\bibinfo  {journal} {Phys. Rev. D}\ }\textbf {\bibinfo {volume}
  {94}},\ \bibinfo {pages} {104020} (\bibinfo {year} {2016})},\ \Eprint
  {https://arxiv.org/abs/1506.06153} {arXiv:1506.06153 [gr-qc]} \BibitemShut
  {NoStop}%
\bibitem [{\citenamefont {{Ruchlin}}\ \emph {et~al.}(2017)\citenamefont
  {{Ruchlin}}, \citenamefont {{Healy}}, \citenamefont {{Lousto}},\ and\
  \citenamefont {{Zlochower}}}]{RucHLZ17}%
  \BibitemOpen
  \bibfield  {author} {\bibinfo {author} {\bibfnamefont {I.}~\bibnamefont
  {{Ruchlin}}}, \bibinfo {author} {\bibfnamefont {J.}~\bibnamefont {{Healy}}},
  \bibinfo {author} {\bibfnamefont {C.~O.}\ \bibnamefont {{Lousto}}},\ and\
  \bibinfo {author} {\bibfnamefont {Y.}~\bibnamefont {{Zlochower}}},\
  }\bibfield  {title} {\bibinfo {title} {{Puncture initial data for black-hole
  binaries with high spins and high boosts}},\ }\href
  {https://doi.org/10.1103/PhysRevD.95.024033} {\bibfield  {journal} {\bibinfo
  {journal} {\prd}\ }\textbf {\bibinfo {volume} {95}},\ \bibinfo {eid} {024033}
  (\bibinfo {year} {2017})}\BibitemShut {NoStop}%
\end{thebibliography}%
\end{document}